\documentclass[11pt]{article} 

\usepackage[utf8]{inputenc} 
\usepackage[toc, page]{appendix}
\usepackage{amsfonts}
\usepackage{bm}
\usepackage{amsmath}
\usepackage{bbm}
\numberwithin{equation}{section}
\numberwithin{figure}{section}
\numberwithin{table}{section}
\usepackage{makecell}

\usepackage[ruled,vlined, linesnumbered]{algorithm2e}

\usepackage[total={6.5in, 8in}]{geometry} 
\usepackage{graphicx} 
\usepackage{float}
\usepackage{caption}
\usepackage{subcaption}
\usepackage{rotating}

\usepackage{booktabs} 
\usepackage{array} 
\usepackage{paralist} 
\usepackage{verbatim} 
\usepackage{hyperref}
\hypersetup{pdftex,colorlinks=true,allcolors=blue}
\usepackage{hypcap}
\usepackage{amsthm}

\usepackage{fancyhdr} 
\usepackage{sectsty}
\allsectionsfont{\sffamily\mdseries\upshape} 

\usepackage[nottoc,notlof,notlot]{tocbibind} 
\usepackage[titles,subfigure]{tocloft} 

\usepackage[font=small,labelfont=bf,tableposition=top]{caption}

\usepackage{authblk}
\usepackage[export]{adjustbox}
\usepackage{mathtools}

\theoremstyle{definition}

\newcommand{\PreserveBackslash}[1]{\let\temp=\\#1\let\\=\temp}
\newcolumntype{C}[1]{>{\PreserveBackslash\centering}p{#1}}
\newcolumntype{R}[1]{>{\PreserveBackslash\raggedleft}p{#1}}
\newcolumntype{L}[1]{>{\PreserveBackslash\raggedright}p{#1}}

\DeclareCaptionLabelFormat{andtable}{#1~#2  \&  \tablename~\thetable}

\providecommand{\keywords}[1]
{
  \small	 
  \textbf{\textit{Keywords---}} #1
}

\title{On the universality of the volatility formation process: \\
when machine learning and rough volatility agree
\thanks{
  The authors gratefully acknowledge financial support from the chair ``Machine Learning \& Systematic Methods in Finance''.
}
}

\author{
  Mathieu Rosenbaum $^1$
  \qquad 
  Jianfei Zhang $^{1, 2}$
}%

\date{%
    $^1$ \footnotesize Ecole Polytechnique, CMAP, 91128 Palaiseau Cedex, France \\%
    $^2$ \footnotesize Exoduspoint Capital Management, 32 Boulevard Haussmann, 75009 Paris, France \\[2ex]%
\today}
\begin{document}
\maketitle

\begin{abstract}
  We train an LSTM network based on a pooled dataset made of hundreds of liquid stocks aiming to forecast the next daily 
  realized volatility for all stocks. Showing the consistent outperformance of this universal LSTM relative to other asset-specific 
  parametric models, 
  we uncover nonparametric evidences of a universal volatility formation mechanism across assets relating past market realizations, 
  including daily returns and volatilities, to current volatilities.
  A parsimonious parametric forecasting device combining the rough fractional stochastic volatility and quadratic rough Heston models with fixed parameters 
  results in the same level of performance as the universal LSTM, which confirms the universality of the volatility formation process from a parametric perspective.   
\end{abstract}

\section{Introduction}
The rough volatility paradigm, where the volatility dynamic is modelled with a fractional
Brownian motion (fBm) with Hurst parameter of order $0.1$, enables us to generate the stylized facts of realized volatilities 
\cite{bennedsen2016decoupling, gatheral2018volatility}. It is also very effective for derivatives pricing and hedging, 
see for example \cite{bayer2016pricing, el2019roughening,el2019characteristic, euch2018perfect}. 
Under the rough fractional stochastic volatility (RFSV) model, one can predict the next realized volatilities following a simple formula 
that requires essentially only one parameter, namely the Hurst parameter \cite{gatheral2018volatility}\footnote{More precisely, only the Hurst parameter is required for 
forecasting log-volatility. When volatility forecasting is concerned, an additional parameter is needed, whose impact is marginal as we will show in 
Section \ref{sec:rfsv_qrt}.}. As it is observed that the 
Hurst parameter is around $0.1$ for a wide range of financial assets \cite{bennedsen2016decoupling, bolko2020roughness, gatheral2018volatility,wu2022rough}, this actually results in a quasi-universal forecasting formula, 
suggesting some universality of the volatility formation process across assets. 
The recently introduced quadratic rough Heston (QRH) model can reproduce at the same time the market observations of SPX and VIX options \cite{gatheral2020quadratic,rosenbaum2021deep}, 
which is known to be challenging with continuous stochastic volatility models. 
A key element to achieving this is the encoding of the feedback effect of price trends on volatility, 
or \textit{strong} Zumbach effect as initially suggested in \cite{zumbach2010volatility}. This motivates us to presume that the potentially 
universal volatility formation mechanism involves also past returns. 

\vskip 0.2in
\noindent
In this work, we aim at confirming this universality of volatility formation mechanism relating past volatilities and returns to current volatilities 
across hundreds of liquid stocks, \textit{i.e.} the values of the involved parameters do not show significant differences among stocks.  
We are not suggesting that the volatility processes of different assets follow exactly the same genesis. The assets with 
different exposures to exogenous events such as earnings announcements will certainly show different volatility distributions.
Here we focus on the mechanism of self-reflexivity of volatility processes. 
As the degree of endogeneity of financial markets is high \cite{filimonov2012quantifying, filimonov2015apparent, hardiman2013critical}, 
we expect that a large part of current volatility levels can be explained by past volatilities and returns. One can imagine that this 
endogenous link between current volatilities and past market realizations emerges from some primary activities at the microstructural scale as suggested in 
\cite{blanc2017quadratic, dandapani2021quadratic, el2018microstructural, jusselin2020no}, leading to a universal volatility formation process across assets.

\vskip 0.2in
\noindent
The universality of the dynamics of stocks is considered in \cite{blanc2014fine,chicheportiche2014fine} in the context of general quadratic autoregressive models, 
which are calibrated with return data, relying on a combination of the generalized method of moments and the maximum-likelihood estimation. 
In this work the access to intraday prices allows us to compute daily realized volatilities.
Then the impact of past volatilities and returns on current volatilities can be inspected by fitting a volatility 
forecasting device dependent on histories of market realizations, by minimizing the error of forecast. 
The heterogeneous autoregressive (HAR) model introduced in \cite{corsi2009simple} is probably 
the most referenced one in this category. It offers a highly parsimonious way to reproduce the slow decay behavior of volatility autocorrelation. 
It is inspired by the heterogeneity and seasonality of market participants' activities, with a volatility prediction typically depending 
on past realized volatilities over one day, one week, and one month. Several extensions have been developed to reproduce additional features 
such as the leverage effect, see for example \cite{corsi2012discrete, patton2015good}. Both the HAR and RFSV models presume some specific formulations 
for the effect of past market realizations on current volatilities. However, we prefer to be agnostic about the dynamic of the volatility formation 
process in the first place. Providing this dynamic is universal across assets, we can fit a forecasting device on the dataset mixing different assets 
to neutralize any asset-dependent effect and reduce the risk of overfitting, which should help the device uncover the ``true'' universal volatility formation process. 
   
\vskip 0.2in
\noindent
Because of our will to be as model-free as possible, in the first part of our work, we choose to apply a nonparametric approach with machine learning. More particularly, we assess the volatility formation 
mechanism by trying to forecast the next daily realized volatility with a long short-term memory (LSTM) network, see \cite{hochreiter1997long}, based on past realized volatilities and returns. 
It is a natural choice to use LSTM other than multilayer perceptron or convolutional neural network for time series learning tasks given its recurrent structure.
Our network is trained with a pooled dataset made of over 800 liquid stocks. We will see that this universal LSTM can consistently outperform the asset-specific parametric
models. We conduct further tests to confirm the targeted universality and its stationarity through time.     

\vskip 0.2in
\noindent
Our universal LSTM network can be viewed as a benchmark for volatility forecasting, 
capturing the universal volatility formation mechanism with a fully nonparametric approach. 
The question now is whether we can uncover this complex ``black box'' 
with a parsimonious parametric forecasting device, while similar performance can be hold. Moreover, 
given the universality in question, we expect that one fixed set of parameter values can apply to all stocks.
The RFSV and QRH models give parametric views about the impacts of past volatilities and returns on current
volatility respectively, as introduced before. For each of them, we conduct comparison tests to confirm that a 
universal model with fixed parameters performs very similarly to those calibrated stock by stock. 
Then we obtain a simple combination of the two aiming at capturing the effect of both past volatilities and returns.
Surprisingly, we show that we can apply this parsimonious forecasting device for all stocks using the same parameters,
with the same level of performance as that of 
the LSTM obtained. This recalls the universality of the volatility formation process 
found by the universal LSTM network, while this time from a parametric perspective.

\vskip 0.2in
\noindent
Applying machine learning methods for volatility forecasting is not something new. LASSO is used in \cite{audrino2020impact} to investigate the impact of
sentiment data on realized volatility. The authors in \cite{rahimikia2020machine} apply LSTM on a rich set of predictor variables collected from 
news and limit order book data. Graph neural networks are studied in \cite{chen2021multivariate} to take cross-asset information into account. 
Machine learning is used in \cite{julien2021} to underline the importance of past returns for VIX forecasting. In our work, past returns will also play important
roles. More precisely, we develop a highly performant volatility forecasting device based on volatility and return data. We highlight the universal nature of the
volatility formation process first from a nonparametric perspective, and then disclose it with a parsimonious parametric formulation. 
In cases with more predictor variables, more complex forecasting methods can be investigated by taking this fundamental feature into account. 

\vskip 0.2in
\noindent
The paper is organized as follows. In Section \ref{sec:model}, we recall the forecasting formulas of HAR and RFSV models.
We also consider the simple autoregressive (AR) model for reasons of comparison. The architecture of our LSTM network is then 
introduced and some key training hyperparameters are listed. We describe in Section \ref{sec:data} the dataset at hand
and the evaluation metrics used for model assessment. The out-of-sample performances of all 
the models are then compared in Section \ref{sec:num_res}. We conduct further tests to confirm the universality of the volatility formation process across assets. 
In Section \ref{sec:rfsv_qrt}, we evaluate the universal versions of the forecasting devices given by the RFSV and QRH models respectively,
and combine them trying to approximate the performance of the above LSTM network.
Finally, we conclude with our main findings in Section \ref{sec:conc}.

\section{Description of the forecasting devices}
\label{sec:model}
\subsection{Parametric methods}
\label{subsec:para_model}
As the ``true'' volatility is unobservable, in our tests we use daily realized volatilities, which we will define in the next section, to inspect the 
volatility formation mechanism.     
We recall first the forecasting formulas of several parametric realized volatility models. In the AR and HAR model, 
the predicted volatility $\hat{\sigma}_t$ is given by a linear combination of past volatilities $(\sigma_{t-p})_{p=1,2,\cdots}$. More specifically, we have: 
\begin{itemize}
  \item AR(p): \\
      $$
        \hat{\sigma}_{t} = \alpha_0 + \sum_{j=1}^{p}\beta_j\sigma_{t-j} \, ,
      $$
      
  \item HAR:
      $$
        \hat{\sigma}_{t} = \alpha_0 + \beta_1\sigma_{t-1} + \beta_2\frac{1}{5}\sum_{j=1}^5\sigma_{t-j} + \beta_3\frac{1}{22}\sum_{j=1}^{22}\sigma_{t-j} \, ,
      $$
\end{itemize}
where $\alpha_0, \beta_{\cdot} \in \mathbb{R}$ and can be estimated via ordinary least squares. We evaluate AR(22) in our tests to make it dependent on the same volatility histories 
as the HAR model. 
\vskip 0.2in
\noindent
In the RFSV model, the logarithm of volatility is modelled by a fBm $W_t^H$ associated with the Hurst parameter $H$: 
\begin{equation}
   \text{d}\log\sigma_t = \nu \text{d}W_t^H \, ,
   \label{eq:fBm_def}
\end{equation}
with $\nu >0$. For a fBm, we have the following property: 
\begin{equation}
  \mathbb{E}[|W^H_{t+\Delta} - W^H_t|^q] = \mathbb{E}[|Z|^q]\Delta^{qH}, \quad Z\sim N(0,1)\, ,
  \label{eq:rough_mom}
\end{equation}  
for any $t\in\mathbb{R}$, $\Delta \geq 0$ and $q>0$. For simplicity, we use the method suggested in \cite{gatheral2018volatility} to estimate $H$ in this work. 
Of course other methods can be tested trying to get more accurate estimations, see for example \cite{bolko2020roughness,wu2022rough}. Here we compute
$$
  m_q(l) = \frac{1}{T-l}\sum_{t=1}^{T-l}|\log\sigma_{t+l}-\log\sigma_{t}|^q \, ,
$$
over intervals of $l\in \{1, 2, \cdots\}$, where $T$ is the number of data points available. By (\ref{eq:rough_mom}), we have
\begin{equation*}
  m_q(l)\simeq \nu^ql^{qH} \, .
\end{equation*}
Then we can regress $\log(m_q(l))$ against $\log(l)$ to estimate $H$ and $\nu$. As proposed in \cite{gatheral2018volatility}, 
we then have the following predictors for $H < \frac{1}{2}$:
\begin{equation}
  \begin{split}
    \widehat{{\log\sigma_{t}}}  &= \frac{\cos(H\pi)}{\pi}\int_{-\infty}^{t-1}\frac{\log\sigma_s}{(t-s+1)(t-s)^{H+1/2}}\text{d}s\, ,\\
    \hat{\sigma}_{t} &= c\exp(\widehat{\log\sigma_{t}}) \, ,
  \end{split}
  \label{eq:rfsv_forecast}
\end{equation}
with 
\begin{equation}
  c = \exp(\frac{\Gamma(3/2 - H)}{2\Gamma(H+1/2)\Gamma(2-2H)}\nu^2) \, .
  \label{eq:rfsv_c}
\end{equation}
In practice, the integral in (\ref{eq:rfsv_forecast}) is truncated to some length of history and then approximated through a Riemann sum.
We will see in Section \ref{sec:rfsv_qrt} that with real data the values of $c$ across assets are close to $1$ and show a very small deviation.
The impact of $\nu^2$ on the forecasting is thus marginal. 
   
\subsection{Nonparametric forecasting with LSTM}
\begin{figure}[!h]
  \centering  
  \includegraphics[width=\textwidth]{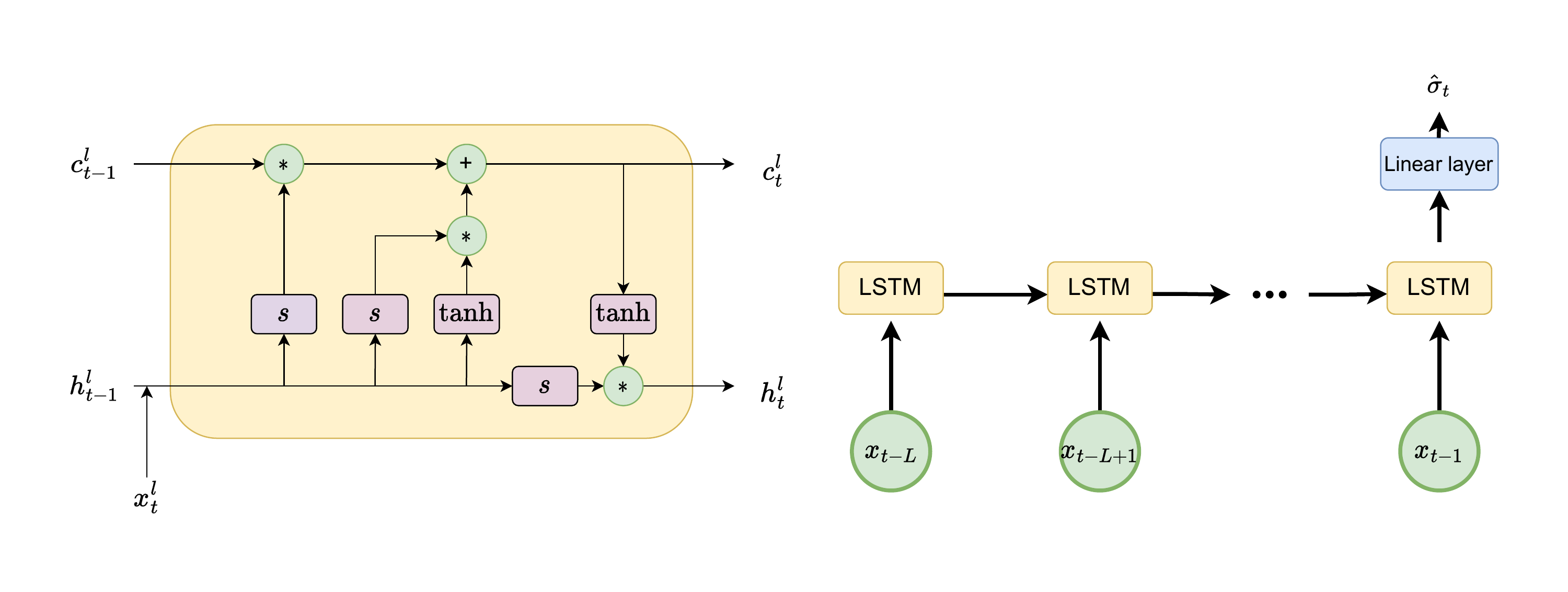}
  \caption{Structure of an LSTM cell (left) and simplified computational graph of the network based on LSTM (right).}
\label{fig:lstm_graph}
\end{figure}
\noindent
A recurrent neural network (RNN) is a type of neural networks suitable for sequential learning tasks, of which LSTM is a popular member. 
The left subfigure of Figure \ref{fig:lstm_graph} illustrates the structure of the layer $l$ of a multilayer LSTM cell. 
For each timestep $t$, the following computations are performed:
\begin{align*}
    i_t^l &= s(W_{ii}^lx_t^l + b_{ii}^l + W_{hi}^lh_{t-1}^l + b_{hi}^l) \, , \\
    f_t^l &= s(W_{if}^lx_t^l + b_{if}^l + W_{hf}^lh_{t-1}^l + b_{hf}^l)  \, ,\\
    g_t^l &= \tanh (W_{ig}^lx_t^l + b_{ig}^l + W_{hg}^lh_{t-1}^l + b_{hg}^l)  \, ,\\
    o_t^l &= s(W_{io}^lx_t^l + b_{io}^l + W_{ho}^lh_{t-1}^l + b_{ho}^l) \, ,\\
    c_t^l &= f_t^l\ast c_{t-1}^l + i_t^l\ast g_t^l \, ,\\
    h_t^l &= o_t^l \ast \tanh(c_t^l) \, ,
\end{align*}
where the superscript indexes the layer, $c_t^l, h_t^l$ are the cell and hidden states, 
$x_t^l$ equals the $t$-th element of the input sequence when $l=1$ or $h_t^{l-1}$ for $l>1$, 
$s(\cdot)$ is the sigmoid function, $\ast$ is the
Hadamard product, $W^l_{\cdot}$ and $b^l_{\cdot}$ are parameters to learn from data.
The main idea of LSTM is to introduce the gates $i, f$ and $o$ to capture long-range dependence in the sequential data and to 
avoid the vanishing gradient problem of classical RNNs \cite{goodfellow2016deep, hochreiter1997long}.
The right subfigure of Figure \ref{fig:lstm_graph} gives the computational graph for predicting the next volatility with the most recent 
$L$ histories. In the following, we test first with $x_t=(\sigma^2_t)$ and then with $x_t=(\sigma^2_t, r_t)$, where $r_t$ is the daily return at time $t$.
The choice of $L$ is given afterward. 
Table \ref{tab:net_summary} summarizes the network architecture and the hyperparameters of training. 
Since in practice neural networks are trained with stochastic gradient descent, to avoid artifacts generated by a particular 
random seed, we fix globally 10 random seeds with which we train 10 independent LSTM networks during each experiment.  
Their average is taken as the final forecast.

\begin{table}[!h]
  \centering
  \begin{tabular}{|m{0.15\textwidth}|m{0.6\textwidth}|}
    \hline
    Architecture & number of layers: 2, dimension of hidden state: 2, activation function: SiLU, output layer: linear \\
    \hline
    Training & optimizer: Adam, loss: mean squared error (MSE), batch size: 512, learning rate: 1e-3, number of epochs: 5 \\
    \hline    
  \end{tabular}
  \caption{Summary of the network architecture and training parameters.}
  \label{tab:net_summary}
\end{table}

\section{Data and evaluation metrics}
\label{sec:data}
\begin{figure}[!h]
  \centering
  \includegraphics[width=.8\textwidth]{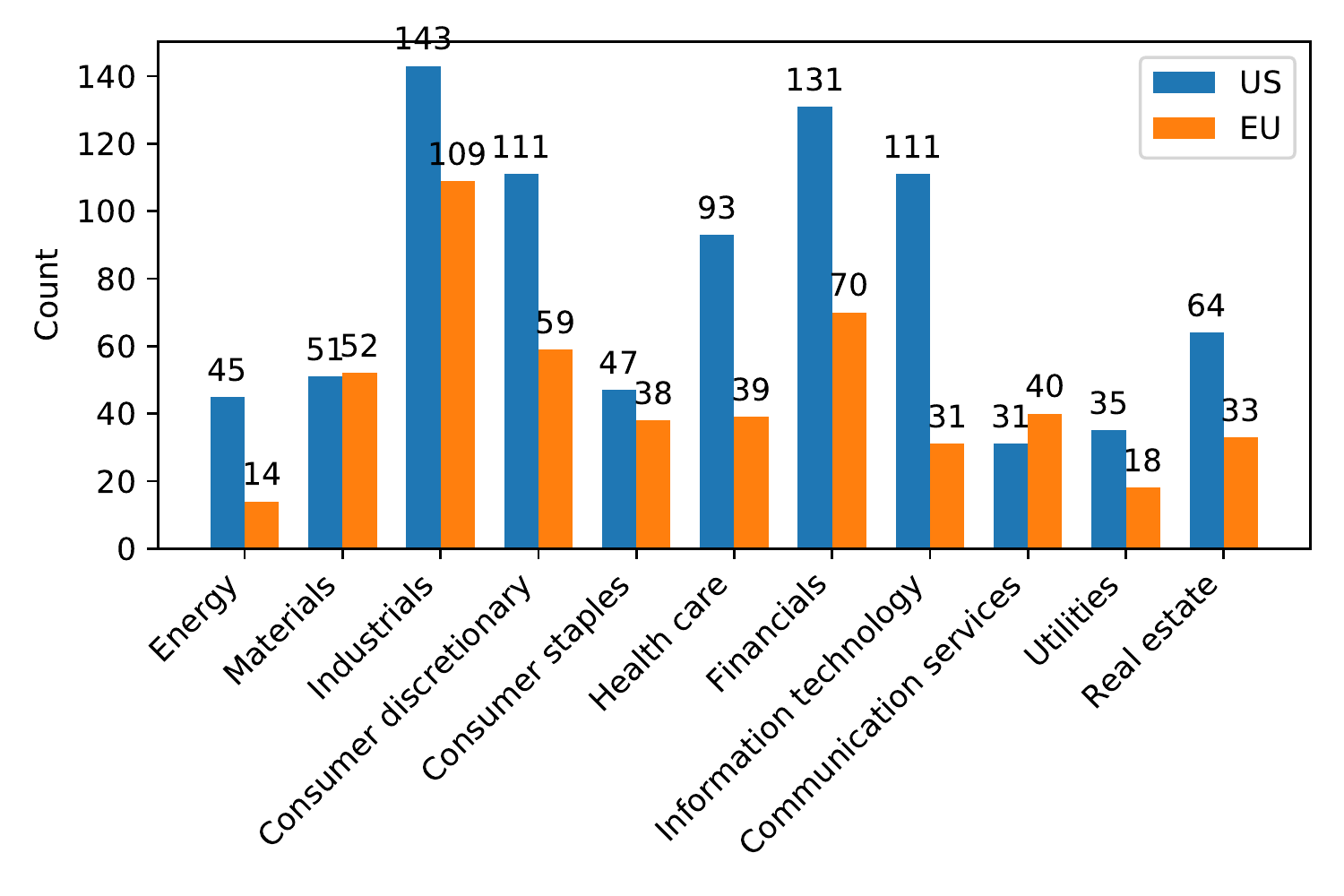}
  \caption{Count of stocks by GICS Sector.}
  \label{fig:count_sector}
\end{figure}

Our dataset contains 5-minutes intraday prices of Russell 1000 and STOXX Europe 600 constituents, for years between 2010 and 2020. 
After removing the ones with considerable missing or abnormal data, 
we get 862 names from the US market and 503 names from the European market. Figure \ref{fig:count_sector} gives their distribution by associated sectors following the Global Industry 
Classification Standard (GICS)\footnote{https://www.msci.com/our-solutions/indexes/gics.}, where we can see a diversity of involved sectors. 
The daily realized volatility is estimated by
\begin{equation*}
  \sigma_t = \sqrt{\sum_i r_{t,i}^2} \, ,
\end{equation*}
where $r_{t,i}$ is defined by the $i$-th 5-minutes intraday return after removing the first and last 30 minutes of the daily trading period. The return of day $t$ is defined as
\begin{equation*}
  r_t = \frac{P_t - P_{t-1}}{P_{t-1}} \, ,
\end{equation*}
where $P_t$ is the closing price of day $t$. To make the data of different stocks have similar scale, 
we perform the following scaling for each stock:
\begin{equation*}
  \centering
  \sigma_t = \frac{\sigma_t}{\sqrt{\langle\sigma^2_t\rangle}}, \qquad r_t = \frac{r_t - \langle r_t \rangle}{\sqrt{\langle(r_t - \langle r_t \rangle)^2\rangle}} \, ,
\end{equation*}
where $\langle\cdot\rangle$ refers to the average over $t$.

\vskip 0.2in
\noindent
We focus mostly on the US market, and the data of the European market is used for an out-of-sample double-check. 
We use the pooled dataset of 862 stocks over years 2010 - 2015 to train the LSTM network. The period 2016 - 2020\footnote{As in this work we focus on the endogenous 
volatility feedback effect, the period 2020-02-01 - 2020-06-01 is excluded from the test set, which is conceived to be largely perturbed by exogenous information.} is used for out-of-sample evaluation.
For the parametric models introduced above, sliding window fit of size 1000 is applied, which means, the model is recalibrated with the last 1000 data points 
for every new forecast.  
We employ MSE as the model evaluation metric, which is defined as follows:
$$
      \text{MSE}(\sigma, \hat{\sigma}) =  \frac{1}{T}\sum_{t=1}^T(\hat{\sigma}_t - \sigma_t)^2  \, ,
$$
where $T$ is the number of trading days of the out-of-sample period. MSE is widely used in forecasting-like tasks. According to \cite{patton2011volatility}, it is a robust and homogeneous metric for model comparison of volatility forecasting.
In this work, we focus more on each model's relative performance
compared to that of the HAR model so that we compute instead 
$(\text{MSE}_{\text{m}}/\text{MSE}_{\text{HAR}})$, 
for $m\in\{\text{AR(22)}, \text{RFSV}, \text{LSTM}\}$.

\section{Capturing universality with LSTM}
\label{sec:num_res}

\begin{figure}[!h]
  \centering
  \includegraphics[width=.7\textwidth]{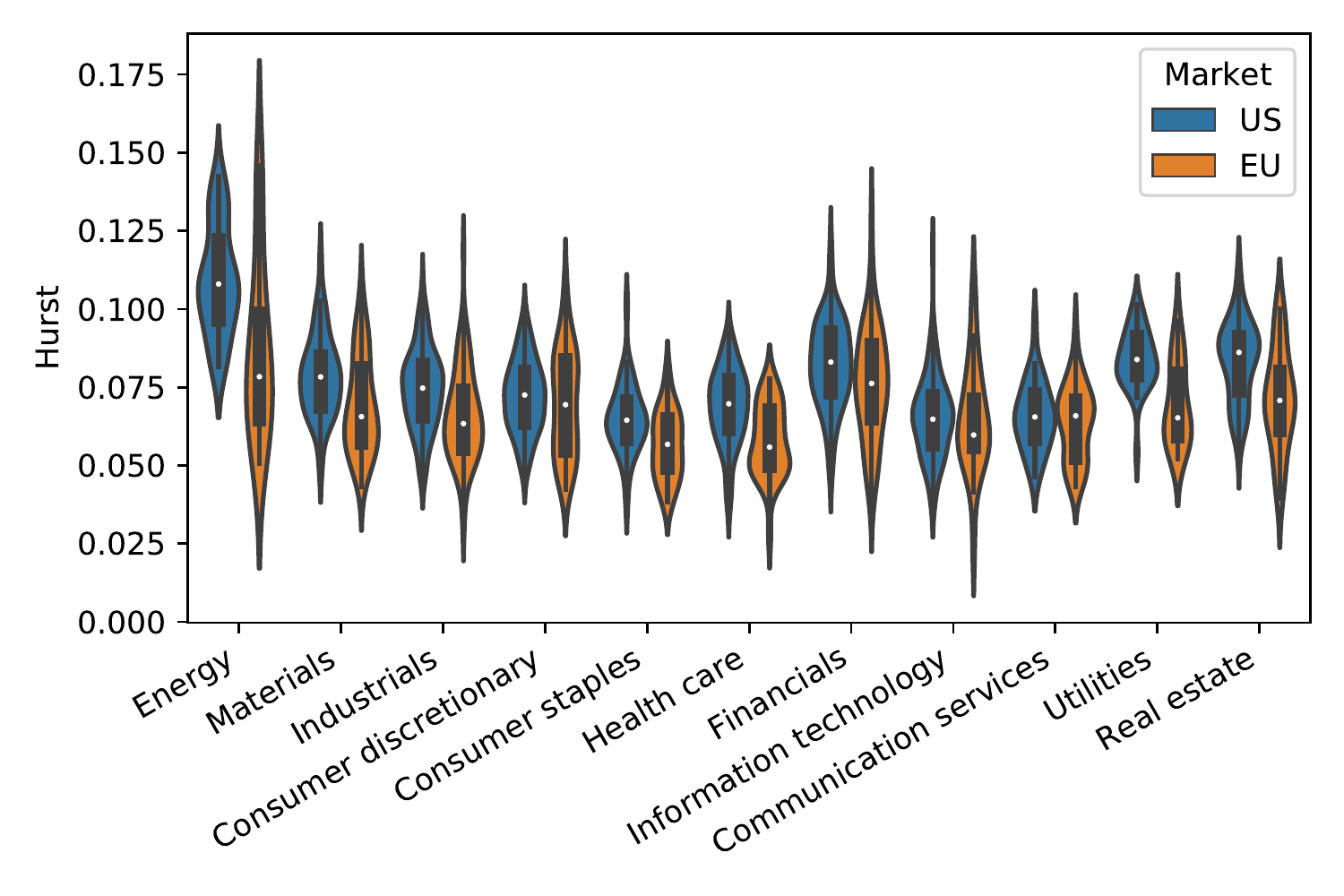}
  \caption{Empirical distribution of the estimated Hurst parameters inside each GICS Sector.}
  \label{fig:hurst_sector}
\end{figure}

We estimate the Hurst parameter of each time series of realized volatility in the dataset following the approach described in Section \ref{subsec:para_model}. 
The resulting distributions by sector shown in Figure \ref{fig:hurst_sector} via violin plot confirm the observations in the literature, 
with $H$ systematically around $0.1$. 
Moreover, we do not remark any significant differences across sectors and markets. 
As presented in the Introduction, this implies a quasi-universal volatility forecasting formula, 
suggesting a universal volatility formation mechanism relating past market realizations to current volatilities. 

\vskip 0.2in
\noindent
We use LSTM to uncover this universality without an \textit{a priori} format. We train first an LSTM network, 
denoted by LSTM$_{var}^{us}$ with $x_t=(\sigma_t^2)$ as input. 
Then we add past returns data, \textit{i.e.} $x_t=(\sigma_t^2,r_t)$, to train another LSTM network LSTM$_{ret}^{us}$. 
We recall that the training is performed with the data of over 800 stocks of the US market. 
In the following, we compare this nonparametric approach with the other parametric methods fitted 
asset by asset in terms of out-of-sample MSE. The universal volatility formation mechanism across assets 
is then investigated through a series of experiments.

\subsection{Parametric vs nonparametric}
\begin{figure}[!h]
  \centering
    \includegraphics[width=.8\linewidth]{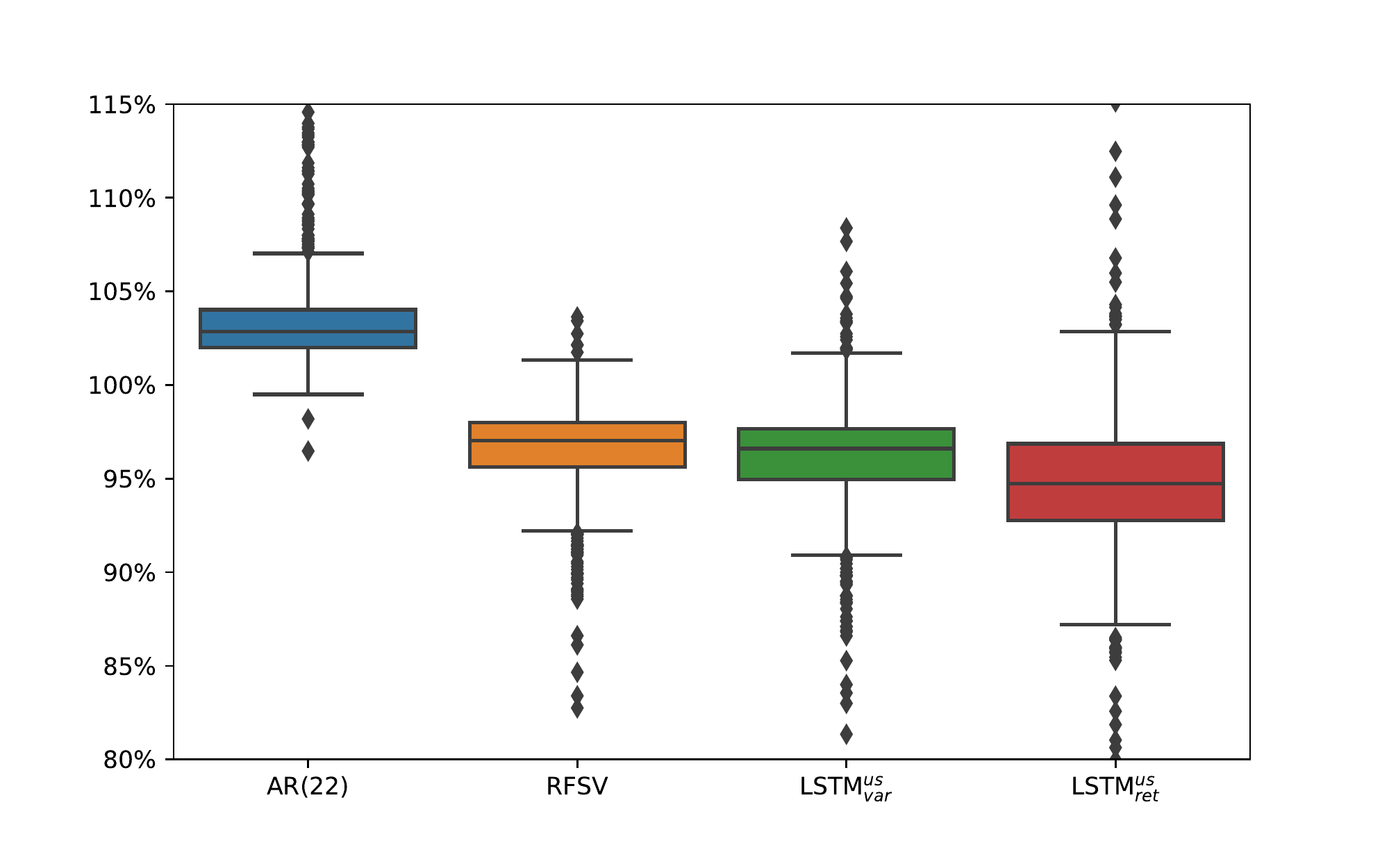}
    \caption{Boxplot showing each model's out-of-sample MSE relative to the HAR model for the stocks of the US market. 
        LSTM$_{var}^{us}$ is trained based on past realized variance, and LSTM$_{ret}^{us}$ uses also past returns.}
  \label{fig:rel_perf_us}
\end{figure}

We recall that the parametric methods introduced before are all fitted on each stock trying to get 
more accurate forecasting, 
while the LSTM models are trained on the pooled dataset of over 800 stocks.
Figure \ref{fig:rel_perf_us} summarizes the out-of-sample performances of the stocks considered. 
We can make the following remarks:
\begin{itemize}
  \item the simple AR(22) model, which imposes no structural constraint on the autoregressive coefficients, 
        underperforms the HAR model. This is understandable as it is more likely
        to overfit,
  \item the RFSV outperforms the HAR model in terms of prediction error.
        It is remarkable as it involves only two free parameters, \textit{i.e.} $\nu$ and $H$ as shown in (\ref{eq:fBm_def}).
        It benefits from the special parametrization of the weights of past volatilities through $H$, 
        as shown in (\ref{eq:rfsv_forecast}), 
  \item the nonparametric models, LSTM$_{var}^{us}$ and LSTM$_{ret}^{us}$ outperform the other parametric models, especially when we incorporate past
        returns data. This indicates that the potential universal volatility formation mechanism across assets, 
        relating past volatilities and returns to current volatilities, allows us to calibrate a universal model 
        based on all assets, where the risk of overfitting is reduced due to enriched realized scenarios. 
\end{itemize}

\begin{figure}[!h]
  \centering
    \includegraphics[width=.8\linewidth]{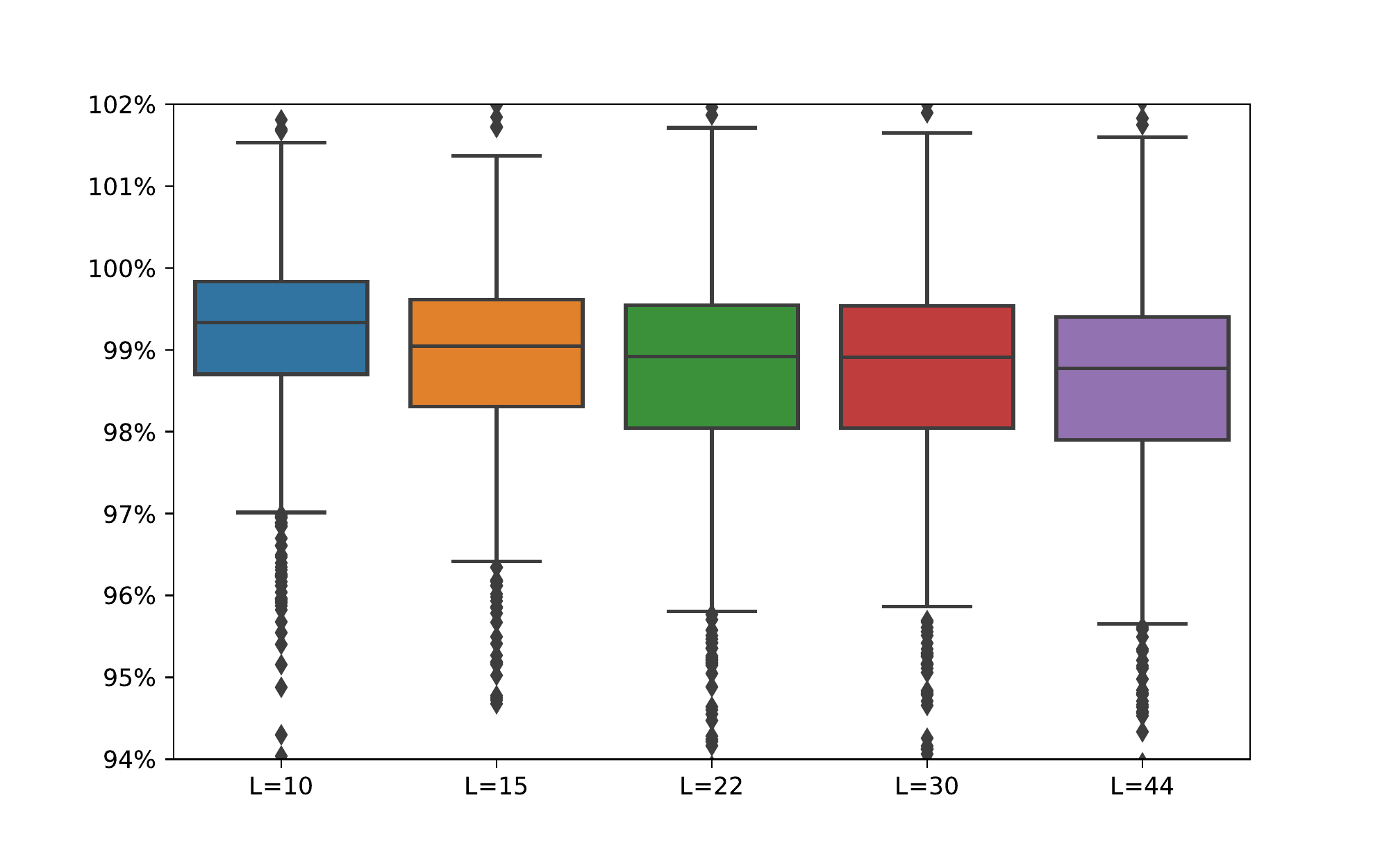}
    \caption{Out-of-sample MSE of LSTM$_{ret}^{us}$ with different $L$ relative to the one with $L=5$.}
  \label{fig:lstm_perf_seqlen}
\end{figure}

\noindent
The length of sequences for training LSTM$_{var}^{us}$ and LSTM$_{ret}^{us}$ is set here to be 22. The scheme 
in Figure \ref{fig:lstm_graph} can adapt to sequences of variable lengths. We choose 22 to make it comparable with 
the HAR model. Figure \ref{fig:lstm_perf_seqlen} shows the forecasting errors of 
LSTM$_{ret}^{us}$ with different values of $L$ relative to that with $L=5$. We can remark that most of valuable information
for the forecast comes from the most recent market realizations over a horizon of fewer than 15 days. The improvement with 
longer input sequences is marginal. We also tested with LSTM networks with more complicated architecture such as higher dimension of
hidden state and more hidden layers, though no significant improvement is observed.    

\subsection{Network inspection}

We see from the above results that a data-driven universal model can outperform asset-specific models.
However, the trained LSTM remains a ``black box''. To shed some light on the universal mechanism captured by the network,
we compute the average sensitivities of its output with respect to the inputs. Then we confirm empirically the existence of some
universal volatility formation mechanism across assets following similar methods as those in \cite{sirignano2019universal}.
Note that when there is no clear indication, the LSTMs in the following are trained with volatility and return data. 

\subsubsection{Local sensitivity visualisation}
\begin{figure}[!h]
  \centering
  \includegraphics[width=\textwidth]{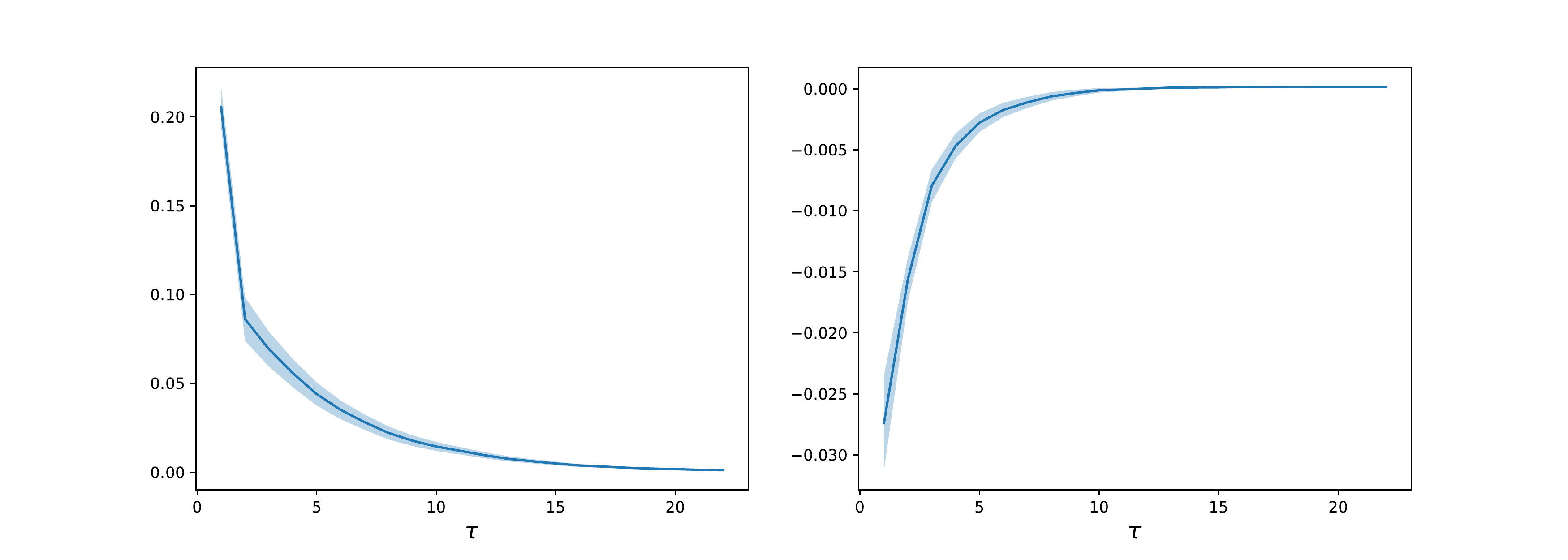}
  \caption{Average $\alpha(\tau)$ (left) and $\beta(\tau)$ (right) across stocks. The shadow regions represent the standard deviation across stocks.}
  \label{fig:local_sensit}
\end{figure}

We define the following local sensitivities:
\begin{equation}
  \alpha_t(\tau) := \frac{\partial \hat{\sigma}_t}{\partial \sigma^2_{t-\tau}} \, , 
  \quad \beta_t(\tau) := \frac{\partial \hat{\sigma}_t}{\partial r_{t-\tau}} \, \quad \text{for} \quad \tau =1,2,\cdots, 22 \, ,
\end{equation}
where $\hat{\sigma}_t$ refers to the output of LSTM$_{ret}^{us}$. Note that these quantities can be computed efficiently with automatic
differentiation. For each stock under study, we evaluate their average over $t$, \textit{i.e.} $\alpha(\tau):=\langle\alpha_t(\tau)\rangle$
and $\beta(\tau):=\langle\beta_t(\tau)\rangle$. Figure \ref{fig:local_sensit} shows the average of $\alpha(\tau)$ and $\beta(\tau)$
across stocks, with their respective standard deviations. The shape of $\alpha(\cdot)$ reflects in particular the well-known slowly decaying autocorrelation function of volatility time series.
The values of $\beta(\cdot)$ are in general one order of magnitude smaller than those of $\alpha(\cdot)$, showing particularly the negative correlation 
between last returns and current volatilities. Interestingly the deviations across stocks of both $\alpha(\cdot)$ and $\beta(\cdot)$ seem to be
very small, implying some universality of the volatility formation mechanism. Of course other partial derivatives of higher orders can also be 
investigated such as $(\frac{\partial^2\hat{\sigma}_t}{\partial r_{t-\tau_1}r_{t-\tau_2}})_{\tau_1,\tau_2=1,\cdots,22}$. However, their magnitude is
relatively smaller and the noise involved in the computation is not negligible. In the following, we confirm this universality in another way, by 
showing mainly that the trained LSTM network can be applied to stocks not included in the training set without significant degradation of performance. 

\subsubsection{Checking universality across assets}
\vskip 0.2in
\noindent
\textbf{- Sector-level universality?}
\vskip 0.15in
\noindent
A natural question is whether this potential universality makes sense at the level of whole market or rather at the level of sector.
One way to verify it is to train sector-specific LSTMs and compare them with the universal LSTM trained with all the sectors. 
For a certain sector-specific LSTM, we are interested in its relative performance to the universal counterpart 
on its covered sector in comparison with the other sectors. If some important sector-dependent volatility formation mechanisms exist,
we expect that this LSTM performs consistently better on its ``local'' sector than the others, where the performance is evaluated 
relative to that of the benchmark - the universal LSTM.
\begin{figure}[!h]
  \centering
  \includegraphics[width=.7\textwidth]{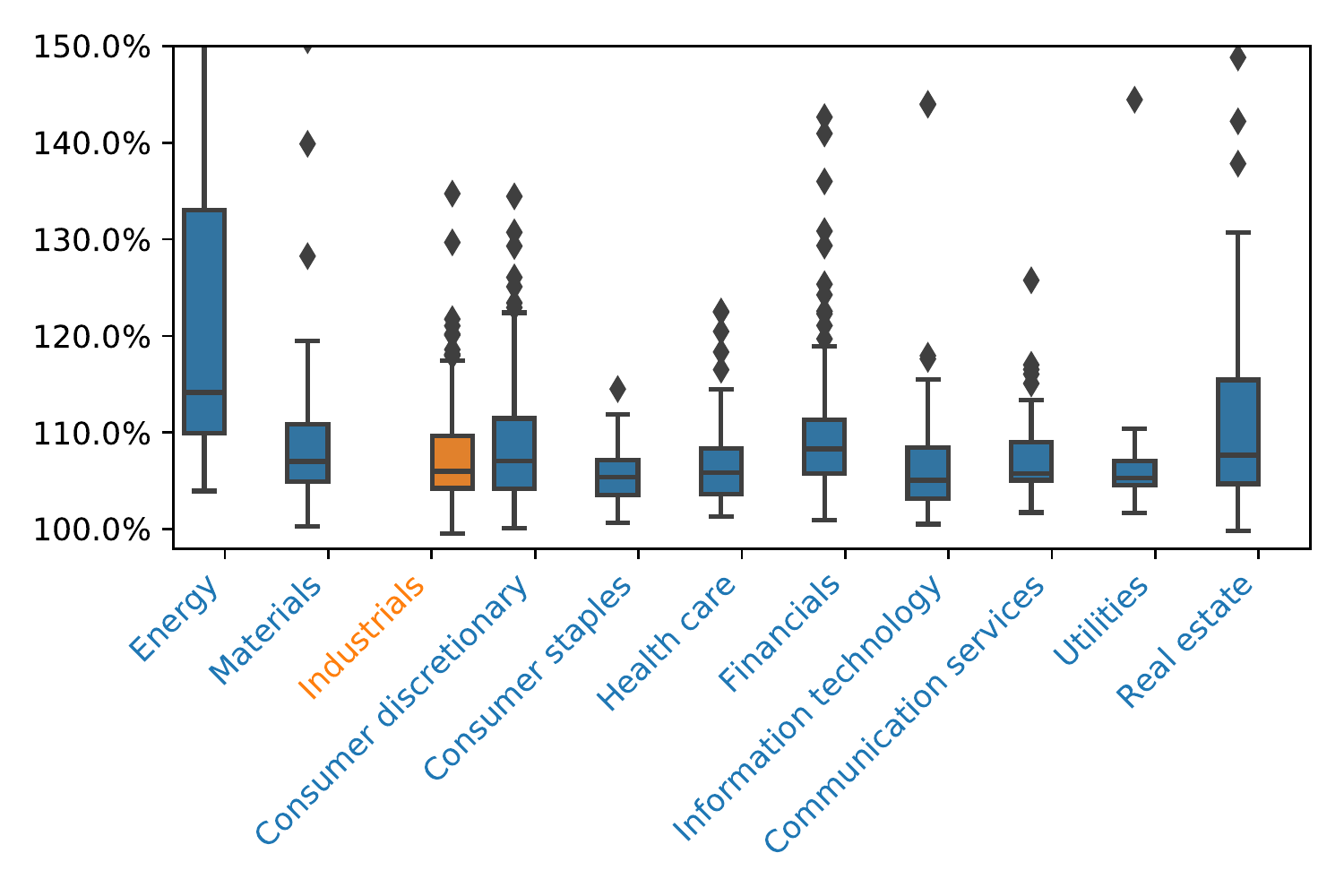}
  \caption{Out-of-sample MSE of the asset-specific LSTM trained with the stocks of the sector \textit{Industrials},
   relative to the universal LSTM covering all sectors.}
  \label{fig:sector_mse_1}
\end{figure}

\begin{figure}[!h]
  \centering
  \includegraphics[width=.7\textwidth]{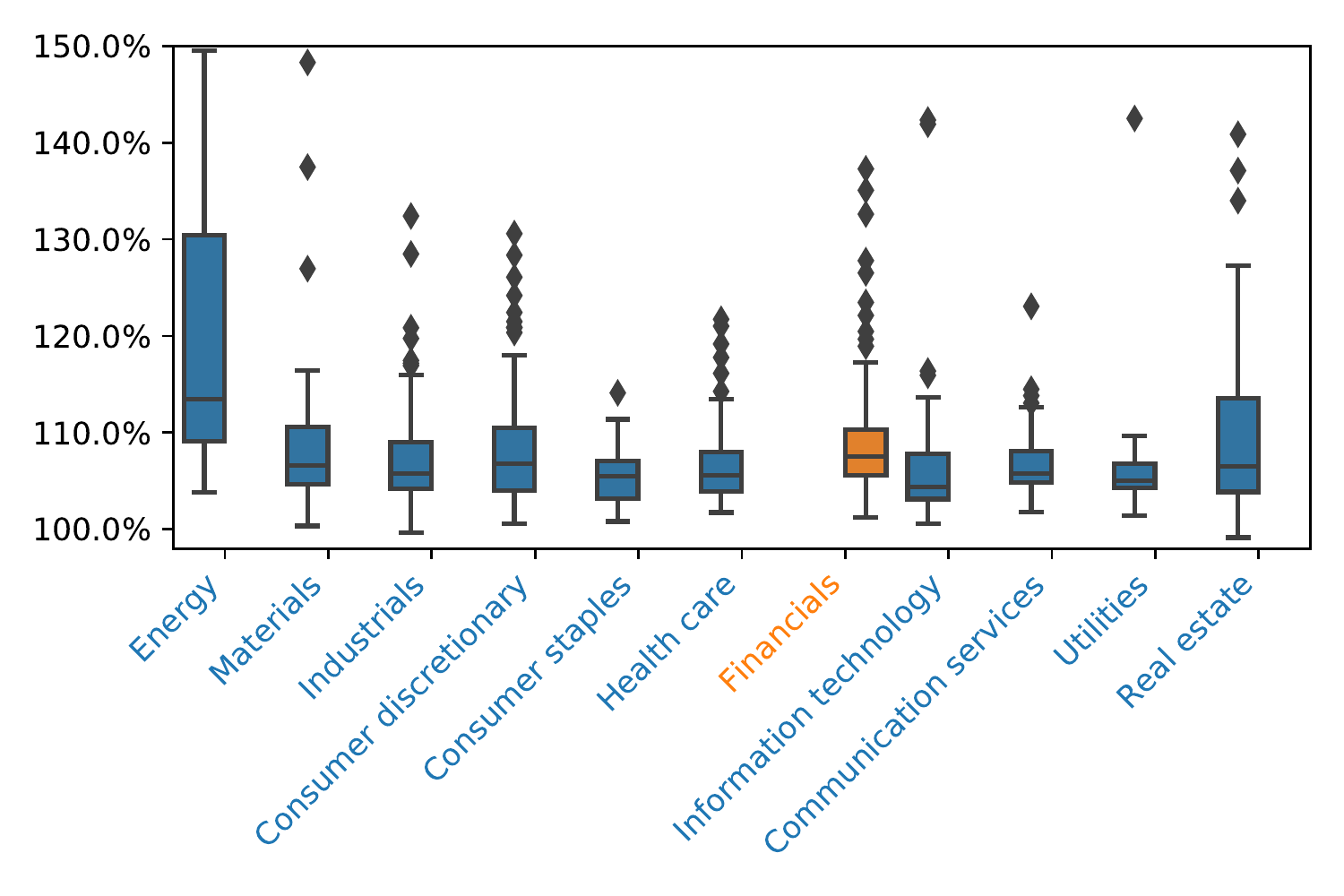}
  \caption{Out-of-sample MSE of the asset-specific LSTM trained with the stocks of the sector \textit{Financials},
    relative to the universal LSTM covering all sectors.}
  \label{fig:sector_mse_2}
\end{figure}

\begin{figure}[!h]
  \centering
  \includegraphics[width=.7\textwidth]{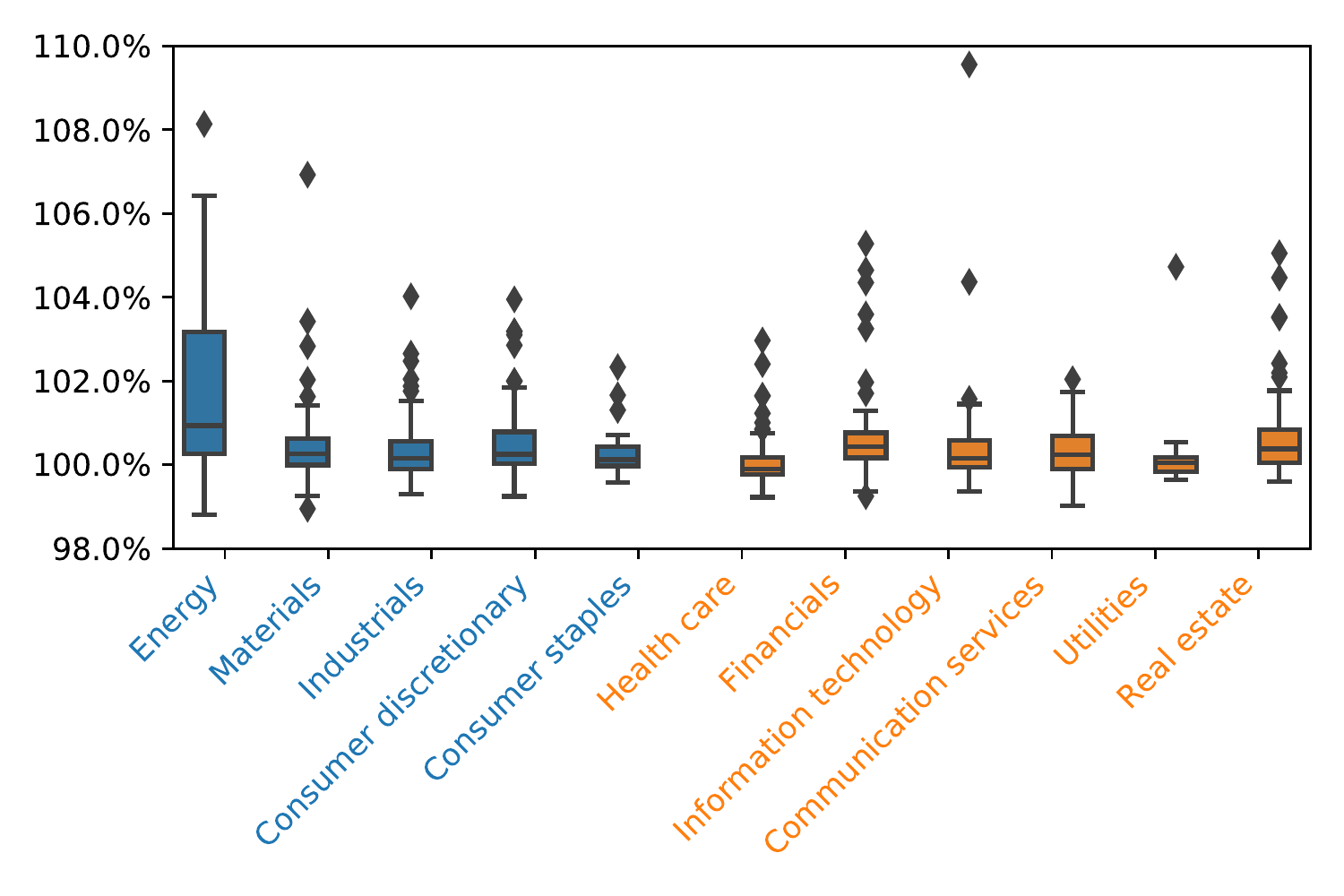}
  \caption{Out-of-sample MSE of an LSTM trained with the stocks of the sectors \textit{Health care}, \textit{Financials}, \textit{Information technology},
          \textit{Communication services}, \textit{Utilities}, and \textit{Real estate}, relative to the universal LSTM.}
  \label{fig:multi_sector_mse}
\end{figure}

\vskip 0.2in
\noindent
Figure \ref{fig:sector_mse_1} and \ref{fig:sector_mse_2} give two examples with respectively the sector \textit{Industrials} and \textit{Financials}. 
Interestingly, in both cases, the sector-specific LSTMs do not show consistently smaller forecasting errors on their covered sectors than the others.
The deterioration of performance on all sectors compared to the universal LSTM is mainly caused by the shrinkage of training data.
Figure \ref{fig:multi_sector_mse} gives an additional example confirming our analysis, where we train an LSTM with multi sectors. 
Thanks to the enriched training set, the forecasting errors of all sectors are universally reduced. The LSTM with only a subset of sectors 
is slightly deteriorated compared to the universal LSTM. These observations conquer the idea of no significant difference on 
the volatility formation mechanism between sectors and support the universality across assets. 
     
\begin{figure}[!h]
  \centering
  \includegraphics[width=\textwidth]{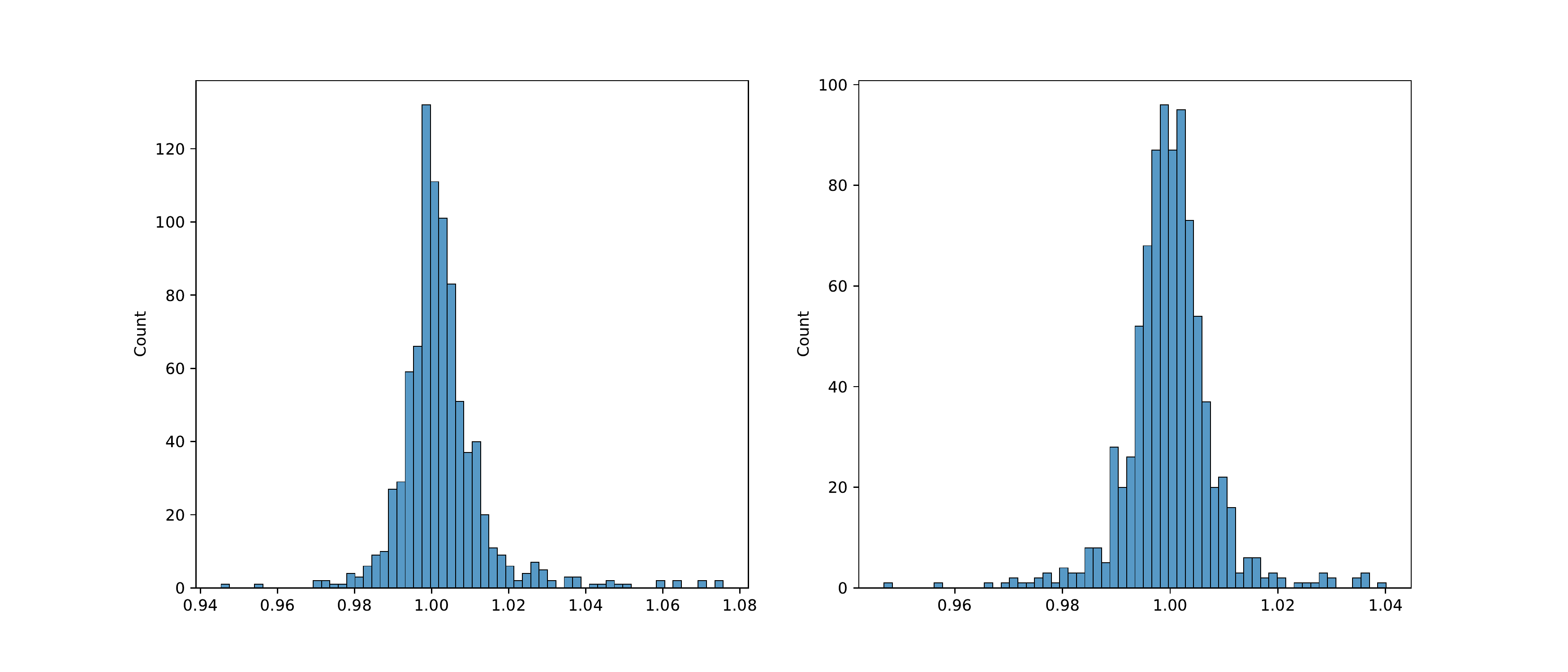}
  \caption{Empirical distribution of out-of-sample MSE of the fine-tuned asset-specific LSTMs relative to the universal LSTM. The left corresponds to 
  the case where all the parameters of the network can be updated. The right is where we only fine-tune the parameters of the output layer.}
  \label{fig:fine_tune}
\end{figure}

\vskip 0.2in
\noindent
\textbf{- Asset-dependent mechanisms?}
\vskip 0.15in
\noindent
The LSTM trained with all the stocks seems to capture the universal link between past market realizations 
and current volatilities. However, we may wonder whether there could be any asset-dependent features that 
the universal network failed to encode. For that,
we fine-tune the universal LSTM on each individual stock to try to add some stock-specific components. 
More specifically, we train an LSTM, whose parameters are initialized with those of LSTM$^{us}_{ret}$, 
with only the data of one specific stock. In the deep learning community, it is also a common practice to fine-tune only the 
output layer of a given pre-trained network.
Here we test both ideas allowing all the parameters or only those of the output layer to be updated. Figure \ref{fig:fine_tune} shows
the out-of-sample performances of these fine-tuned models relative to the original one. We do not remark any significant reduction
of forecasting error after the parameters of the network being ``adapted'' locally to each stock. This seems to reject again any important 
asset-dependent mechanism relating past realizations to current volatilities, at least for those that can be captured nonparametrically by an LSTM network.     

\begin{figure}[!h]
  \centering
  \includegraphics[width=.8\textwidth]{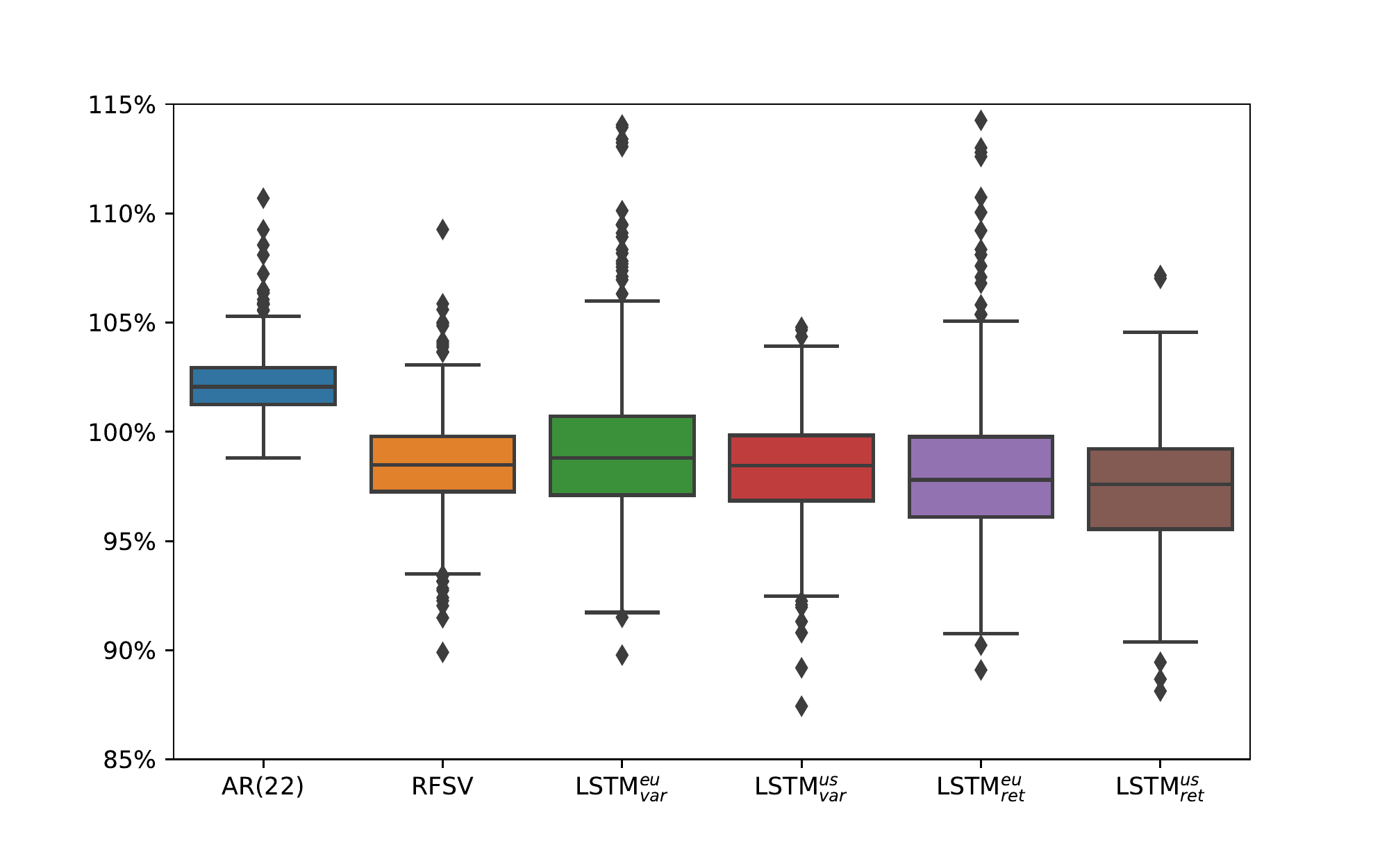}
  \caption{Boxplot showing models’ out-of-sample performances relative to the HAR model for the stocks
  of the European market. Note that LSTM$_{\cdot}^{eu}$ and LSTM$_{\cdot}^{us}$ are trained 
  on stocks of European and US markets respectively.}
  \label{fig:rel_perf_eu}
\end{figure}

\vskip 0.2in
\noindent
\textbf{- Universality across markets?}
\vskip 0.15in
\noindent
The above tests are all done on US market data. Does this universality hold even across markets? 
In other words, can the LSTM based on the US market be directly applied to stocks of the European market without deterioration of performance? 
To see that, with European market data, we fit the considered parametric forecasting devices and train
two LSTM networks, LSTM$_{var}^{eu}$ and LSTM$_{ret}^{eu}$, similarly to the test as shown in Figure \ref{fig:rel_perf_us}. 
Besides, we test also the LSTM$_{var}^{us}$ and LSTM$_{ret}^{us}$, which are trained with US market data,
on the stocks of the European market. From Figure \ref{fig:rel_perf_eu}, the universal models depending on past volatilities and returns, 
\textit{i.e.} LSTM$_{ret}^{eu}$ and LSTM$_{ret}^{us}$, obtain
the best out-of-sample performance. Moreover, LSTM$_{var}^{us}$ and LSTM$_{ret}^{us}$ get slightly better results compared 
to their respective counterparts, LSTM$_{var}^{eu}$ and LSTM$_{ret}^{eu}$. This supports again the existence of a universal volatility formation mechanism across assets, 
which do not differ from one market to another.
LSTM$_{var}^{us}$ and LSTM$_{ret}^{us}$ benefit from the abundance of data of the US market, and thus capture better the endogenous processes relating 
past volatilities and returns to current volatilities.  

\begin{figure}[!h]
  \centering
  \includegraphics[width=.8\linewidth]{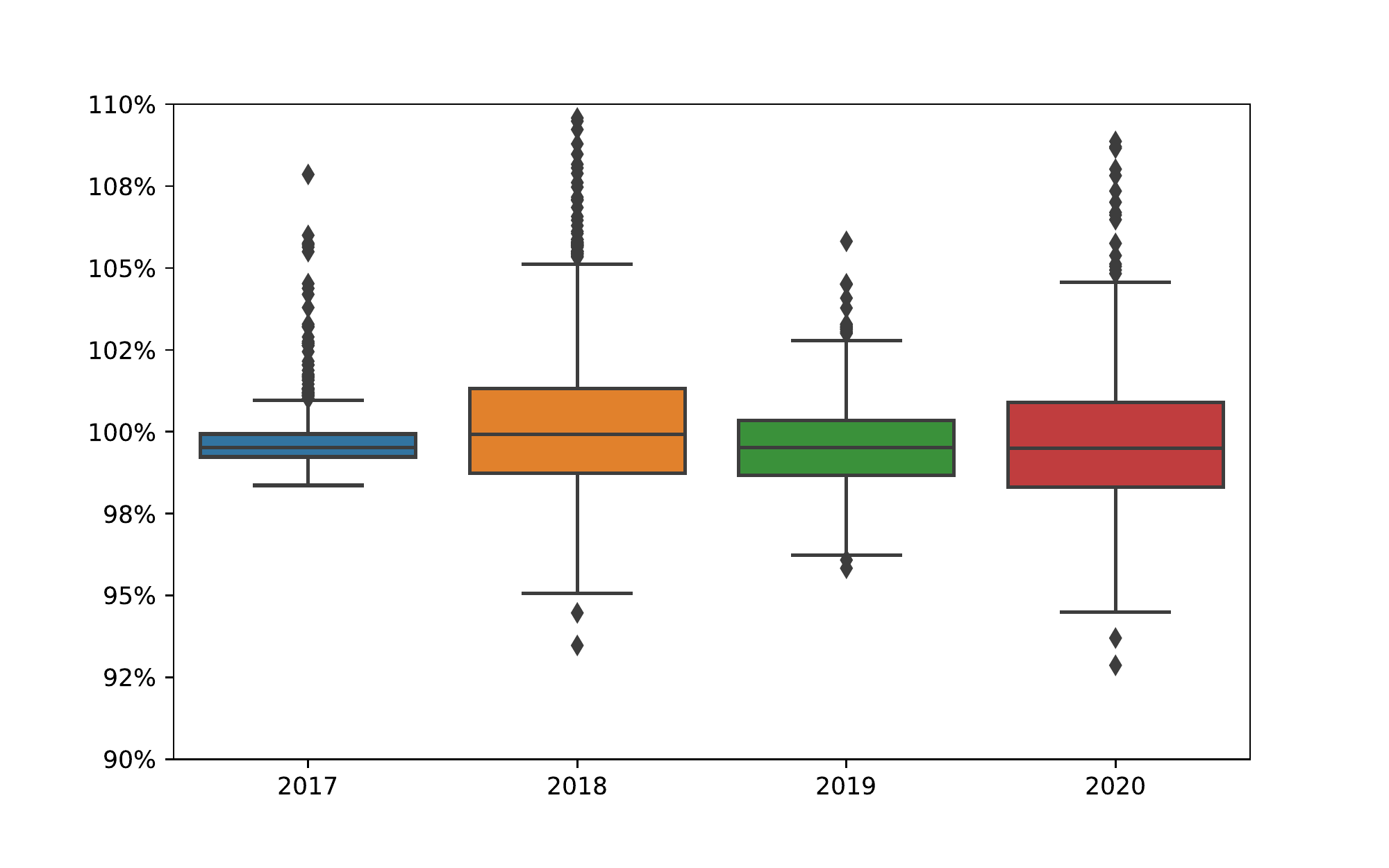}
  \caption{Out-of-sample performance of the ``dynamic'' models relative to the ``static'' model.}
  \label{fig:roll_static_rel_perf}
\end{figure}

\vskip 0.2in
\noindent
\textbf{- Stability across time?}
\vskip 0.15in
\noindent
With financial time series, it is a common practice to recalibrate models based on a sliding window of the most recent data as 
it is perceived that there are changes in market regimes. To see whether the universal volatility formation mechanism is time-varying, we apply the idea of ``dynamic'' prediction. At the end
of each year, we retrain an LSTM based on the data of last six years and use it for the forecasting of the following year. We monitor the out-of-sample performances
of these ``dynamic'' models relative to the initial ``static'' one trained on data of 2010 - 2015. As shown in Figure \ref{fig:roll_static_rel_perf}, 
the ``dynamic'' models perform very closely to the ``static'' one, not showing any superiority on forecasting for the periods just following their training sets. This gives evidence that the universality is also stationary over time. 
One does not need to retrain the universal model frequently trying to adapt to varying market regimes.

\newpage
\section{Uncovering the universal volatility formation process}
\label{sec:rfsv_qrt}

\begin{figure}[!h]
  \centering
  \includegraphics[width=\linewidth]{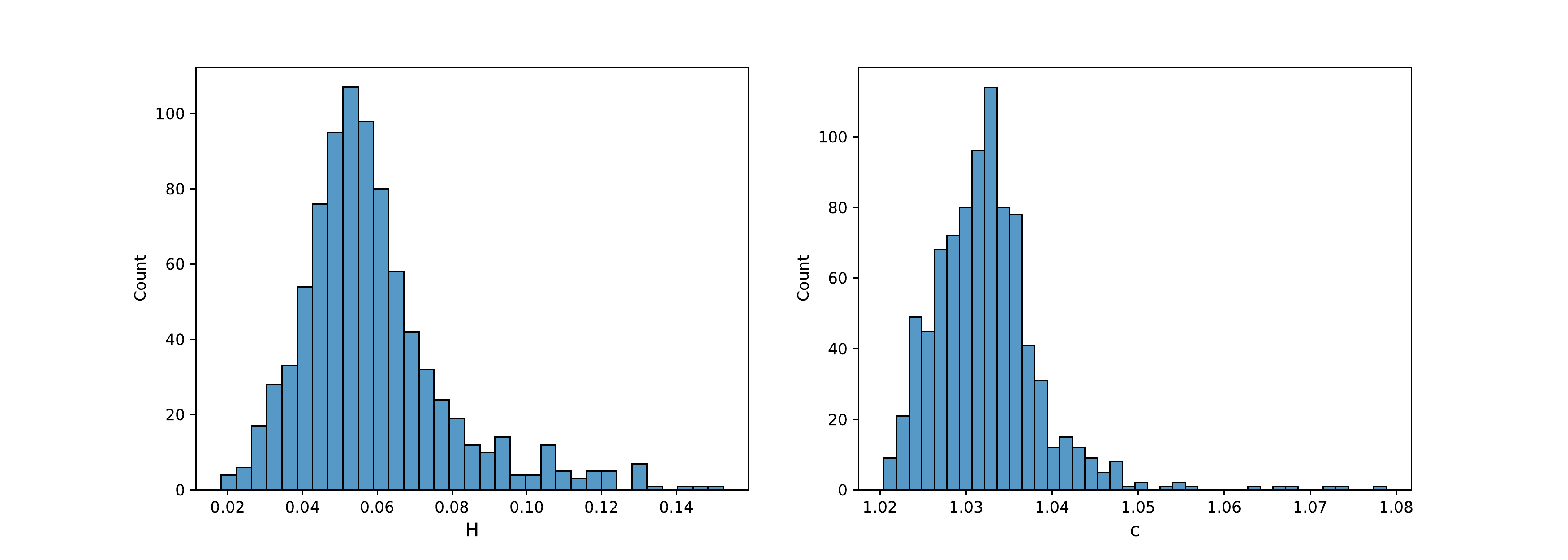}
  \caption{Distribution of $H$ and $c$ calibrated with data of 2012 - 2015 of US market, with medians Median$(H)\sim 0.055$ and Median$(c)\sim 1.03$.}
  \label{fig:hurst_c_dist}
\end{figure}

\begin{figure}[!h]
  \centering
  \begin{subfigure}{.5\textwidth}
    \centering
    \includegraphics[width=\linewidth]{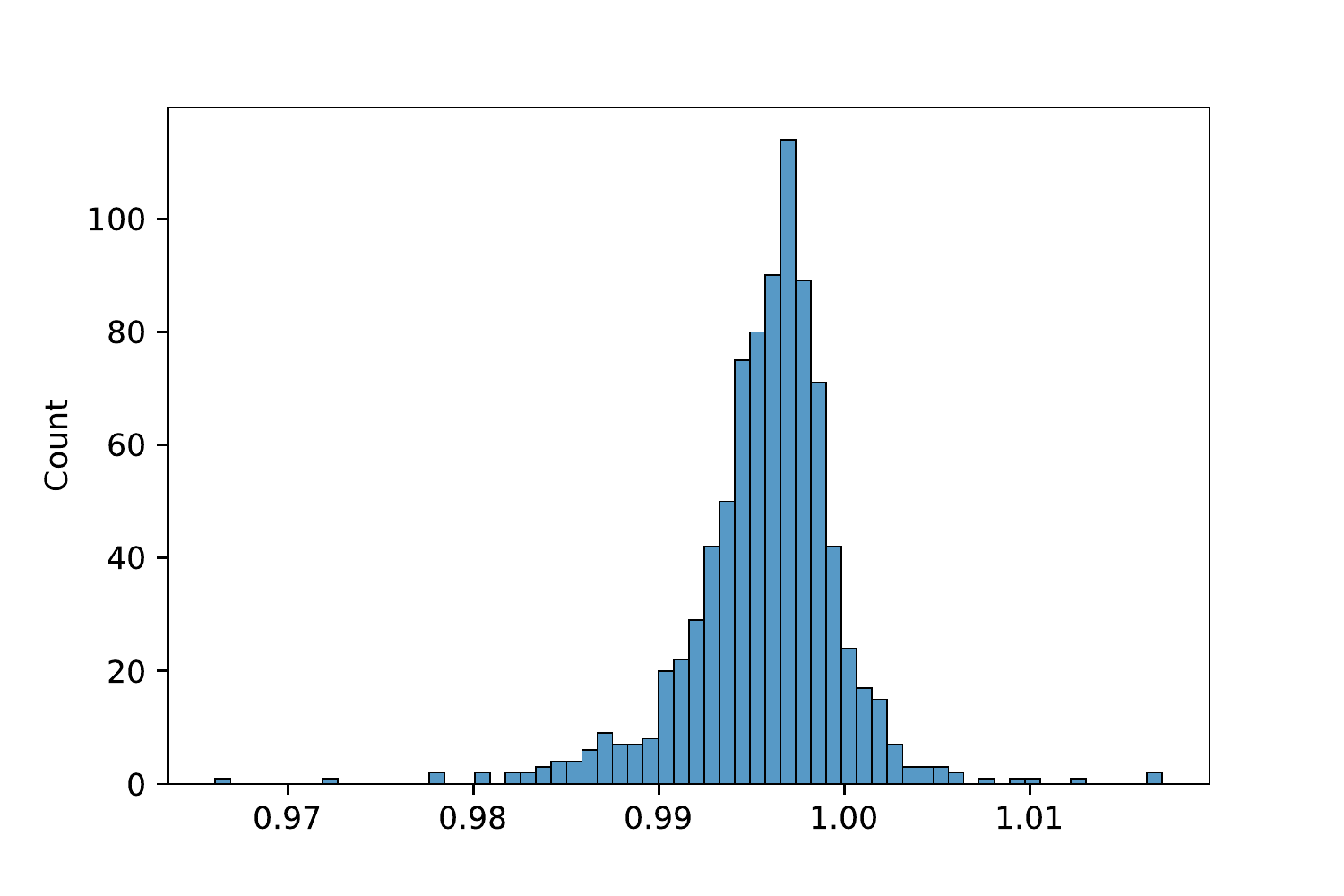}
    \caption{US market}
  \end{subfigure}%
  \begin{subfigure}{.5\textwidth}
    \centering
    \includegraphics[width=\linewidth]{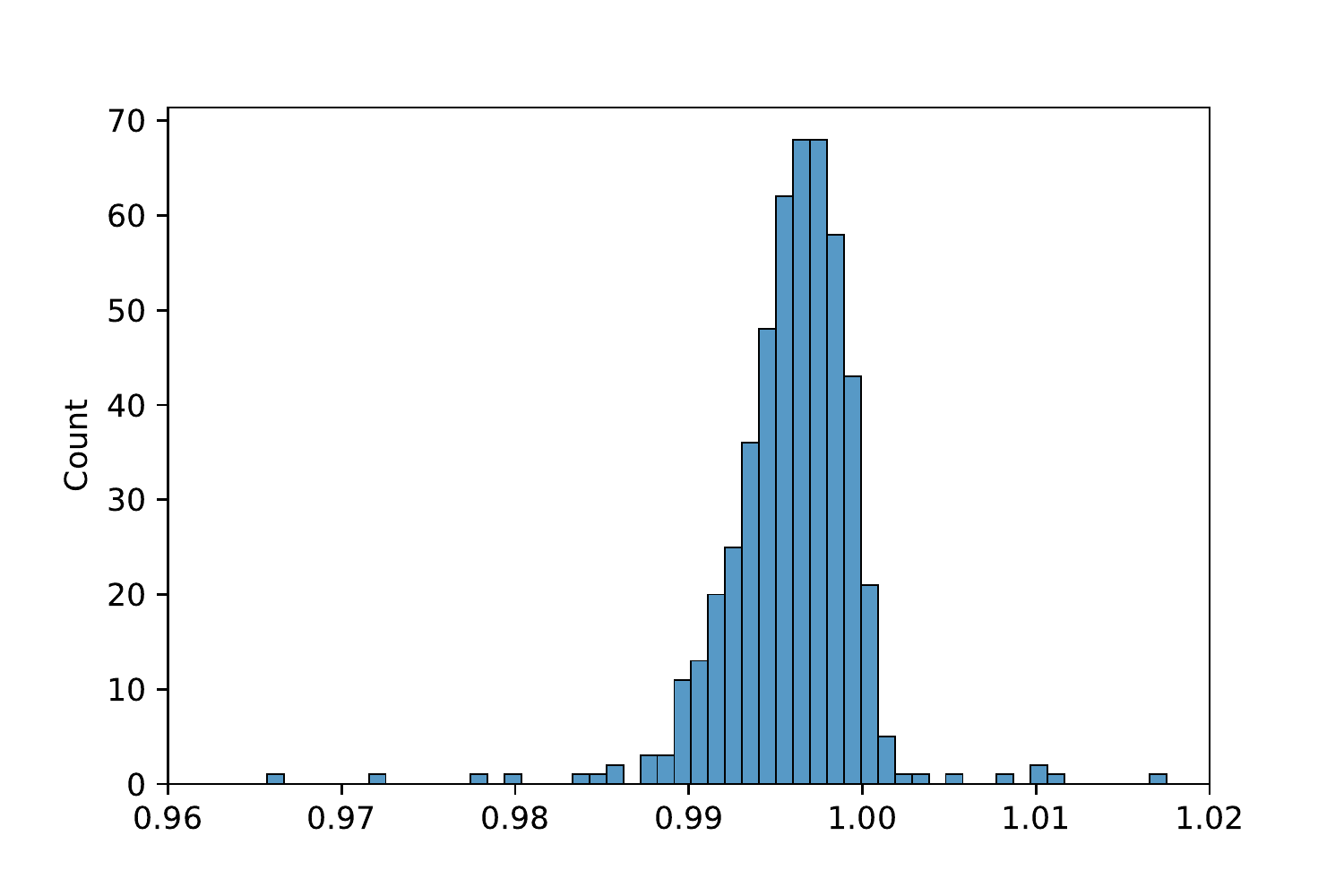}
    \caption{European market}
  \end{subfigure}
  \caption{Out-of-sample MSE of the RFSV forecast with fixed parameters, relative to that of calibration on each stock.}
  \label{fig:rfsv_fix_mse}
\end{figure}

The above results confirm a universal volatility formation mechanism across assets, relating past volatilities and returns to current volatilities,
from a nonparametric perspective. We wonder now whether we can disclose this universal nature with a parsimonious parametric formulation, while keeping
similar level of forecasting performance as that of the LSTM network. 
The RFSV model gives a parametric view of the universal link between past and current volatilities. The results in Figure \ref{fig:rel_perf_us} and 
\ref{fig:rel_perf_eu} shows that the RFSV model can get similar out-of-sample forecasting performance as the LSTMs trained with past volatilities data.  
The recently introduced QRH model encodes \textit{strong} Zumbach effect describing the feedback effect of price trends on volatility, which 
is missing in the RFSV model. In the following, we first verify the universality of both models. For each model, we will see that similar forecasting performance can be obtained 
with properly fixed parameters as in the case where parameters are calibrated on each stock. Then we evaluate a simple combination of these two universal parametric
forecasting devices, comparing its performance with the universal LSTM.
\vskip 0.2in
\noindent
\textbf{- RFSV}
\vskip 0.1in
\noindent
As shown in (\ref{eq:rfsv_forecast}), forecasting volatility in the RFSV model involves two parameters, $H$ and $\nu^2$, where the impact of $\nu^2$ is imposed only through the $c$ 
defined in (\ref{eq:rfsv_c}). 
Figure \ref{fig:hurst_c_dist} shows the estimation results of $H$ and $c$ with data of 2012 - 2015 of US market, by following the method introduced
in Section \ref{subsec:para_model}. The deviations across stocks do not seem to be large for both $H$ and $c$. Other methods can be tested in the future 
trying to reduce the estimation errors. Here for illustrating the idea, we fix $H$ and $c$ with the observed medians, \textit{i.e.} $H=0.055$, $c=1.03$, 
for the forecasting of all the stocks considered, including those of the European market. Figure \ref{fig:rfsv_fix_mse} gives its out-of-sample MSE relative to
those fitted on each stock based on a sliding window of 1000 days. Interestingly, the device with predetermined parameters performs slightly 
better than the asset-specific ones, which could be impacted more by estimation errors due to limited data. 
This suggests the universality of the RFSV method in the sense that the same parameters work for all stocks.  

\vskip 0.2in
\noindent
\textbf{- QRH}
\vskip 0.1in
\noindent
Following the idea on Zumbach effect of the QRH model, we propose the 
following forecasting device:
\begin{align}
  \hat{\sigma}_t^2 &= a(Z_{t-1} - b)^2 + c  \, \nonumber ,\\
  Z_{t} &= \sum_{i=1}^nc_i^dZ_{i,t} \, , \label{eq:Z_sum}\\
  Z_{i,t} &= e^{-\gamma_i^d}Z_{i,t-1} + r_t, \quad Z_{i,0} = z_{i,0}, \quad i=1,\cdots, n \, , \label{eq:z_factor}
\end{align}
where $a,b,c > 0$ and $(c_i^d, \gamma_i^d)_{i=1,\cdots,n}$ are given by the same multi-factor approximation of 
the rough kernel function $K(t):=\frac{t^{H-1/2}}{\Gamma(H+1/2)}$ as that 
used in \cite{rosenbaum2021deep} \footnote{With daily data we use $c_i^d=\frac{c_i}{\sum_i c_i}, \gamma_i^d=\frac{\gamma_i}{250}$, 
where $c_i$ and $\gamma_i$ following the same definition as in \cite{rosenbaum2021deep}}. 
Given some $H$ and the number of factors $n$ for the approximation, we recall that $(c_i^d, \gamma_i^d)_{i=1,\cdots,n}$ are not free parameters 
to calibrate. Equations (\ref{eq:Z_sum}, \ref{eq:z_factor}) can be seen as a multi-factor discretization of the process
$Z_t = \int_{-\infty}^t\frac{(t-s)^{H-\frac{1}{2}}}{\Gamma(H+\frac{1}{2})}\sigma_s\text{d}W_s$\footnote{Applying the multi-factor 
approximation avoids choosing a predetermined $\Delta$ to compute $Z_t \simeq \int_{t-\Delta}^t\frac{(t-s)^{H-\frac{1}{2}}}{\Gamma(H+\frac{1}{2})}\sigma_s\text{d}W_s$. 
Long-range information can be well captured by the one-step update of $(Z_{i,\cdot})_{i=1,\cdots,n}$ following (\ref{eq:z_factor}).}, 
which is essentially a moving average of past returns.
The above model is not very sensitive to $(z_{i,0})_{i=1,\cdots,n}$ because of the exponential decay. In practice given a time series $(r_t, \sigma_t)_{t=1,\cdots,T}$ of 
length $T$, we can take $z_{i,0}=0$ and compute $(Z_{i, t})_{t=1,\cdots,T}$ following (\ref{eq:z_factor}). 
Then we can simply discard the $N$ initial samples $(Z_{i,t}, \sigma_t)_{t=1,\cdots, N}$ from the original data history. 
With the remaining data, we search for the $(a,b,c)$ minimizing the error $\sum_{t=N+1}^T(\hat{\sigma}_t - \sigma_t)^2$ using some optimization 
algorithm\footnote{Or one can simply regress $\sigma_t^2$ 
against $(Z_t, Z_t^2)$.}.

\begin{figure}[!h]
  \centering
  \includegraphics[width=\textwidth]{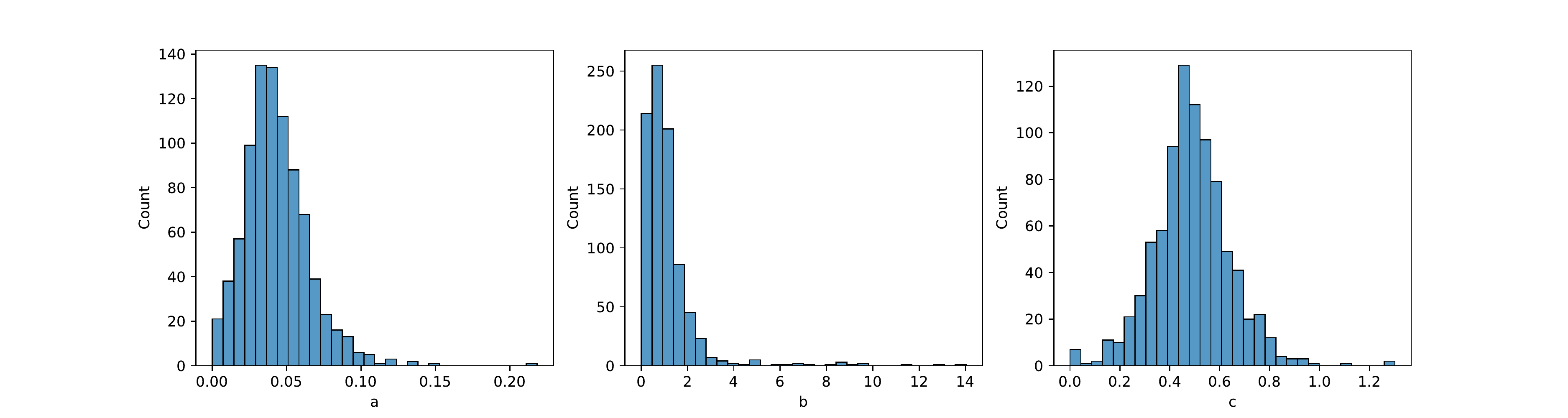}
  \caption{Distribution of $a, b,$ and $c$ calibrated with data of 2012-2015 of US market, with medians Median$(a)\sim 0.043$, Median$(b)\sim 0.74$ and Median$(c)\sim0.55$.}
  \label{fig:qrheston_dist}
\end{figure}

\begin{figure}[!h]
  \centering
  \begin{subfigure}{.5\textwidth}
    \centering
    \includegraphics[width=\linewidth]{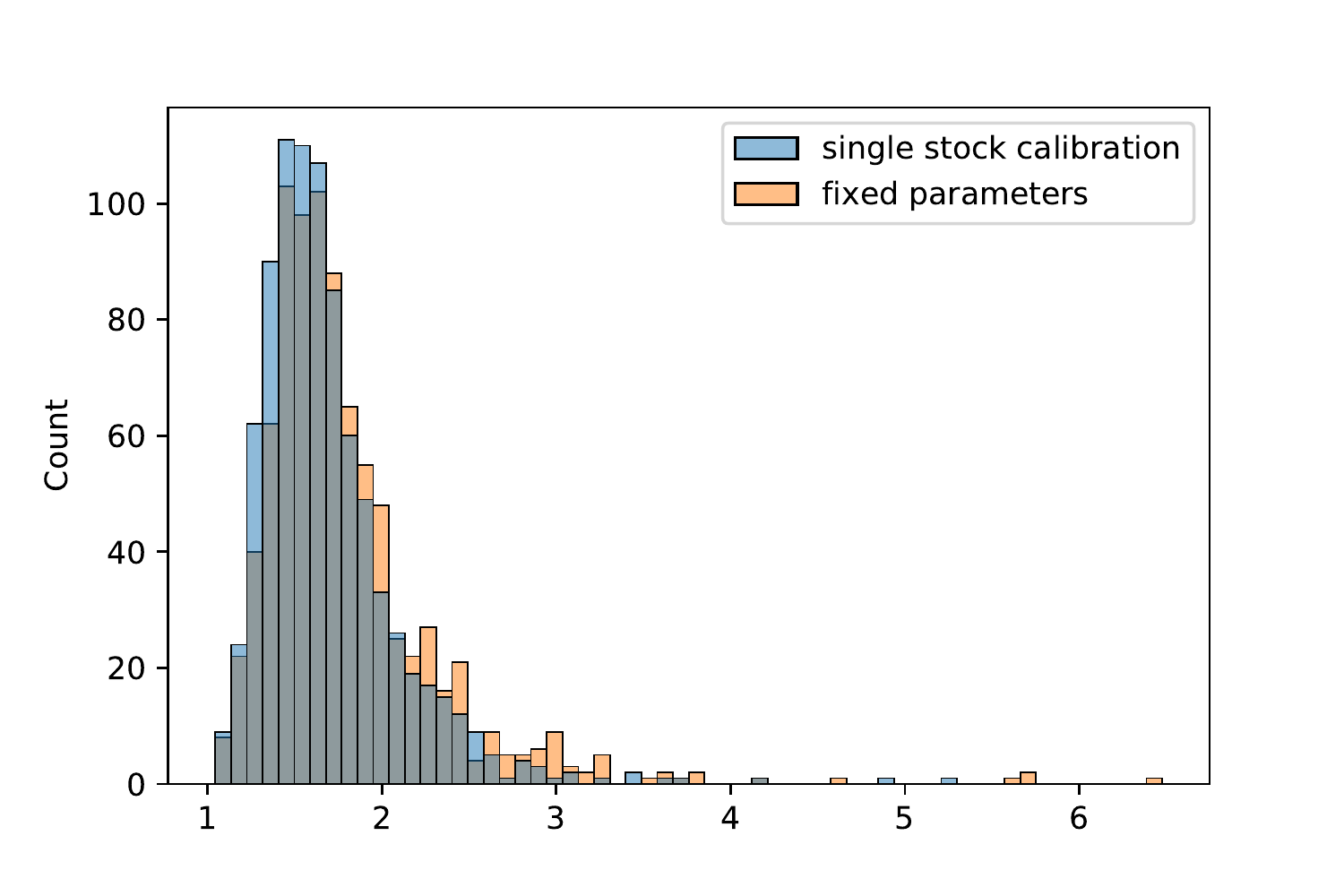}
    \caption{US market}
  \end{subfigure}%
  \begin{subfigure}{.5\textwidth}
    \centering
    \includegraphics[width=\linewidth]{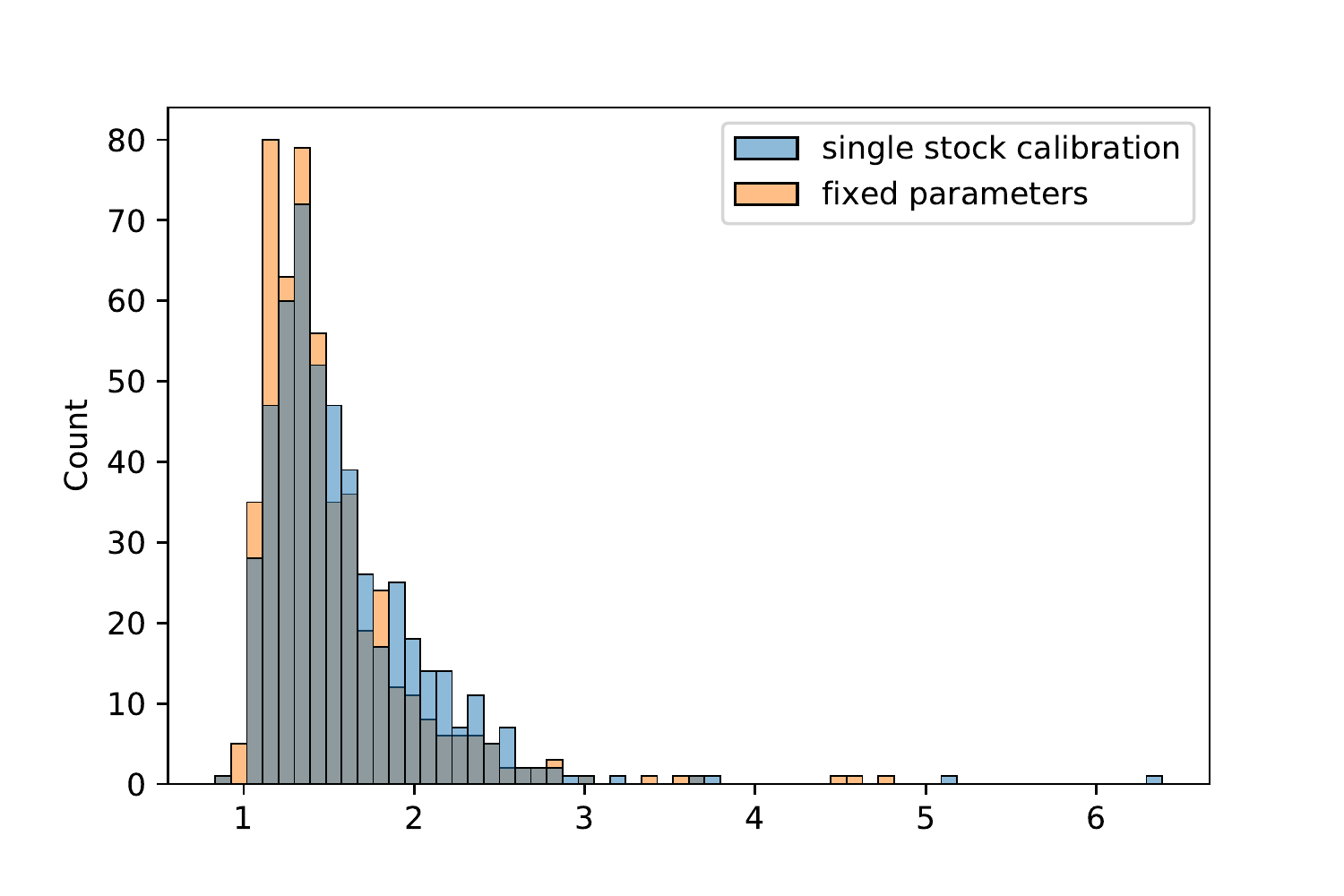}
    \caption{European market}
  \end{subfigure}
  \caption{Out-of-sample MSE of the QRH forecast with calibration on each stock and with fixed parameters, relative to the HAR forecast.}
  \label{fig:rheston_fix_mse}
\end{figure}

\vskip 0.2in
\noindent
Figure \ref{fig:qrheston_dist} shows the resulting distributions of calibrated $a, b$ and $c$ with data of 2012-2015 of US market.
The deviations across stocks seem to be large, especially for $b$. In fact as indicated in Figure \ref{fig:local_sensit}, the impact of past returns on 
current volatilities is not as apparent as that of past volatilities, it is not surprising to get noisy estimations of $(a,b,c)$.
We are interested in how the forecasting performance evolves when we use fixed $(a,b,c)$ for all stocks. Similarly as above, we test with their median observations. 
Figure \ref{fig:rheston_fix_mse} shows its resulting out-of-sample MSE, along with that of stock-specific calibrations where $(a,b,c)$ are fitted on 
each stock based on a sliding window of 1000 days. We do not remark that the forecasting with fixed parameters performs significantly worse than the other.
Note that the QRH forecast is outperformed by the HAR as the former does not use past volatilities.   

\begin{figure}[!h]
  \centering
  \includegraphics[width=.8\textwidth]{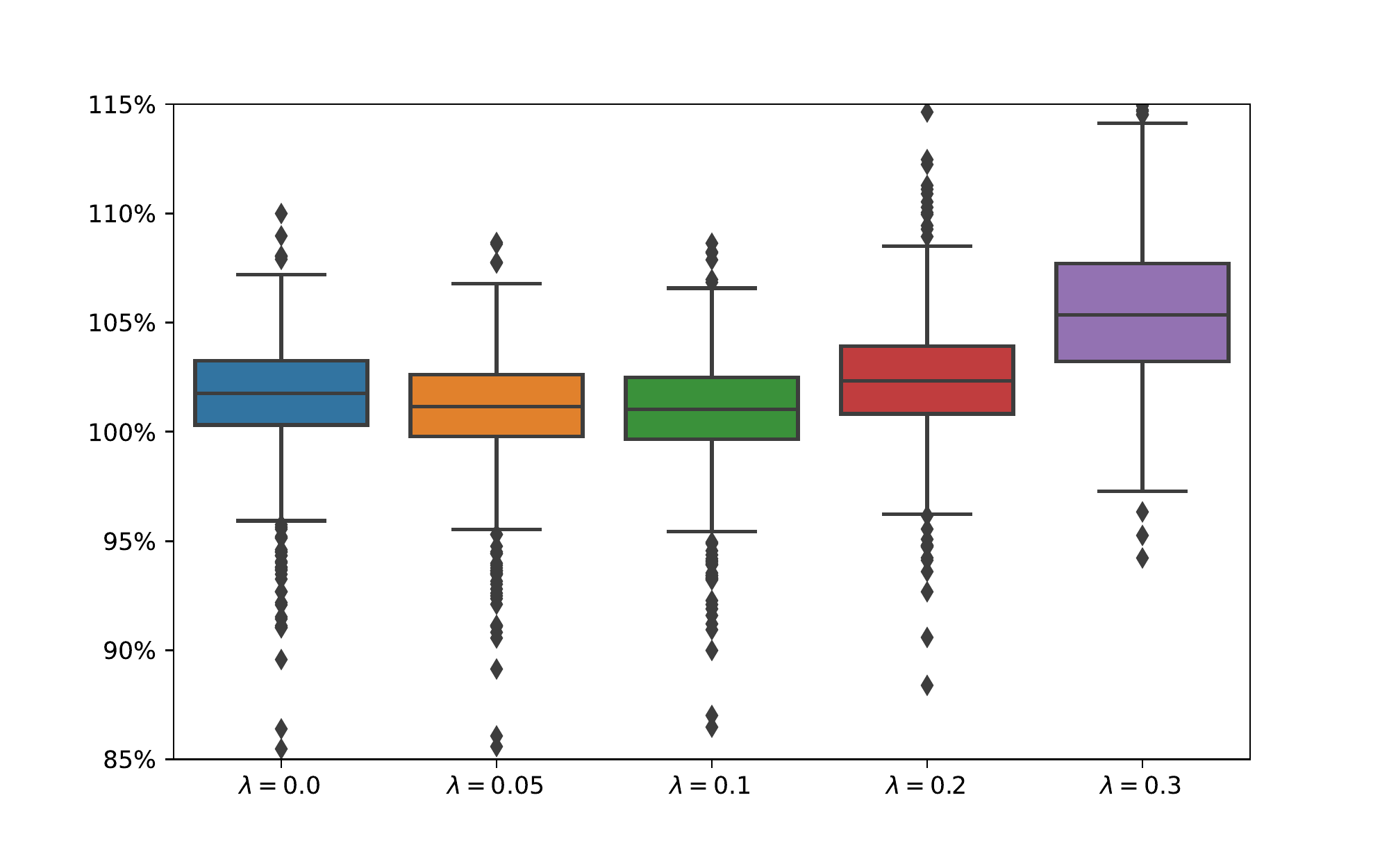}
  \caption{Out-of-sample performance of the forecast $(1-\lambda)\hat{\sigma}^{RFSV} + \lambda\hat{\sigma}^{QRH}$ relative to LSTM$_{ret}^{us}$ in the US market.}
  \label{fig:rfsv_qrh_us}
\end{figure}

\begin{figure}[!h]
  \centering
  \includegraphics[width=.8\textwidth]{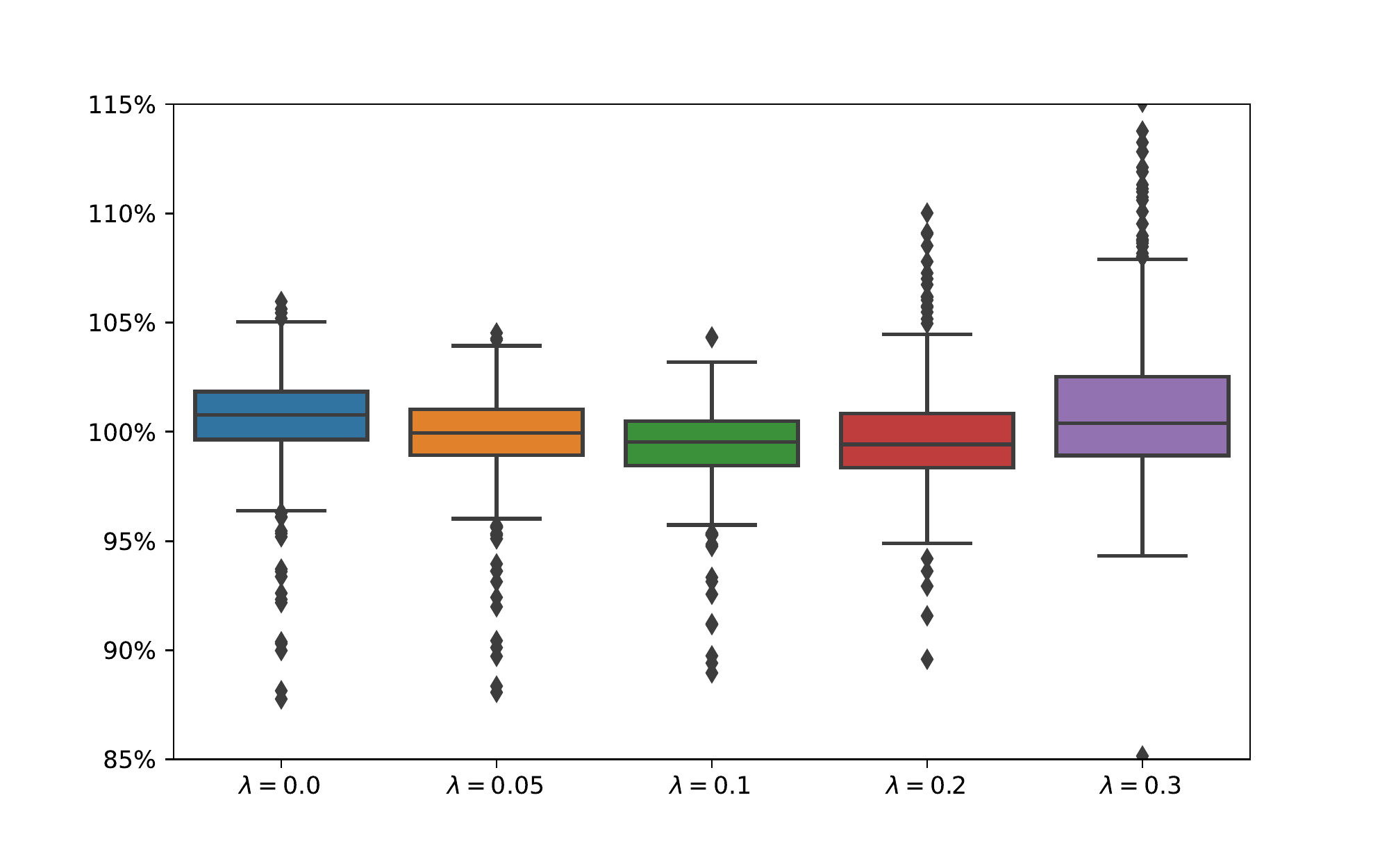}
  \caption{Out-of-sample performance of the forecast $(1-\lambda)\hat{\sigma}^{RFSV} + \lambda\hat{\sigma}^{QRH}$ relative to LSTM$_{ret}^{eu}$ in the European market.}
  \label{fig:rfsv_qrh_eu}  
\end{figure}

\vskip 0.2in
\noindent
\textbf{- RFSV + QRH}
\vskip 0.2in
\noindent
We have seen that for the RFSV and QRH forecasts, fixing properly the parameters can obtain similar performance as allowing the parameters to be fitted on each stock.
Since they do not depend on the same data for the forecast, we use the following combination as the final forecast:
$$
   (1-\lambda)\hat{\sigma}^{RFSV} + \lambda\hat{\sigma}^{QRH} \, ,
$$  
where $\lambda \in (0, 1)$, $\hat{\sigma}^{RFSV}$ and $\hat{\sigma}^{QRH}$ stand for the RFSV and QRH forecasts respectively. We evaluate this mixed forecast for different $\lambda$, 
with the other parameters fixed to the same values as above, \textit{i.e.}
$$
  H = 0.055, c=1.03, a=0.043, b=0.74, c=0.55 \, .
$$
Figure \ref{fig:rfsv_qrh_us} and \ref{fig:rfsv_qrh_eu} show their forecasting performance relative to that of LSTM$_{ret}^{us}$ in US and European markets respectively.
We remark that with $\lambda\sim 0.1$, the blended forecast can benefit from the complementary characteristic given by the RFSV and QRH forecasts, leading to the same level
of performance as the universal nonparametric LSTM$_{ret}^{us}$.   

\noindent
Basically, the future volatility is made of two components depending respectively on past volatilities and 
past price trends. The first one describes the expected future log-volatility from a linear combination of 
past log-volatilities. The associated coefficients depend only on the roughness parameter H. A universal $H$
 works for all stocks. The second component corresponds to the feedback effect of past price trends on 
 volatility. The past price trends are built with a rough kernel, and their impact is expressed in 
 a quadratic form, with essentially constant parameters across assets.  
\section{Conclusion}
\label{sec:conc}
In this work, we shed a light on the universality of the volatility formation process relating past market realizations to current volatilities at the daily scale.
For that, we apply the LSTM network to forecast the next daily volatility, based on past daily volatilities and returns. 
The universal model, trained on a pooled dataset of hundreds of stocks, outperforms consistently the asset-specific 
parametric models based on past volatilities. 
Interestingly, similar superior performances hold on assets that are not part of the training set, 
even on those of a different market. Fine-tuning the universal model with the data of each stock does not help improve
the performance. These observations suggest the existence of a universal volatility formation mechanism from a nonparametric perspective. 
Then we uncover the learned universal volatility formation process with a parsimonious parametric formulation 
involving the RFSV and QRH forecasting devices.
It turns out that a simple combination of the RFSV and QRH forecasts with fixed parameters perform similarly to our LSTM network.
From a parametric perspective, this confirms again the ``targeted'' universality, showing that the main features of this universal volatility
formation process can be well described by 
the rough volatility paradigm boosted with \textit{strong} Zumbach effect.    
Our findings can also be helpful for volatility forecasting with richer predictor variables. For example, one can imagine a mixed forecast, containing the 
endogenous part induced by past market realizations via the uncovered universal mechanism and other temporal effects contributed by some exogenous variables.      

\newpage
\bibliography{ref}

\begin{thebibliography}{10}

\bibitem{audrino2020impact}
F.~Audrino, F.~Sigrist, and D.~Ballinari.
\newblock The impact of sentiment and attention measures on stock market
  volatility.
\newblock {\em International Journal of Forecasting}, 36(2):334--357, 2020.

\bibitem{bayer2016pricing}
C.~Bayer, P.~Friz, and J.~Gatheral.
\newblock Pricing under rough volatility.
\newblock {\em Quantitative Finance}, 16(6):887--904, 2016.

\bibitem{bennedsen2016decoupling}
M.~Bennedsen, A.~Lunde, and M.~S. Pakkanen.
\newblock Decoupling the short-and long-term behavior of stochastic volatility.
\newblock {\em arXiv preprint arXiv:1610.00332}, 2016.

\bibitem{blanc2014fine}
P.~Blanc, R.~Chicheportiche, and J.-P. Bouchaud.
\newblock The fine structure of volatility feedback {II}: Overnight and
  intra-day effects.
\newblock {\em Physica A: Statistical Mechanics and its Applications},
  402:58--75, 2014.

\bibitem{blanc2017quadratic}
P.~Blanc, J.~Donier, and J.-P. Bouchaud.
\newblock Quadratic {H}awkes processes for financial prices.
\newblock {\em Quantitative Finance}, 17(2):171--188, 2017.

\bibitem{bolko2020roughness}
A.~E. Bolko, K.~Christensen, M.~S. Pakkanen, and B.~Veliyev.
\newblock Roughness in spot variance? a {GMM} approach for estimation of
  fractional log-normal stochastic volatility models using realized measures.
\newblock {\em arXiv preprint arXiv:2010.04610}, 2020.

\bibitem{chen2021multivariate}
Q.~Chen and C.-Y. Robert.
\newblock Multivariate realized volatility forecasting with graph neural
  network.
\newblock {\em arXiv preprint arXiv:2112.09015}, 2021.

\bibitem{chicheportiche2014fine}
R.~Chicheportiche and J.-P. Bouchaud.
\newblock The fine-structure of volatility feedback {I}: Multi-scale
  self-reflexivity.
\newblock {\em Physica A: Statistical Mechanics and its Applications},
  410:174--195, 2014.

\bibitem{corsi2009simple}
F.~Corsi.
\newblock A simple approximate long-memory model of realized volatility.
\newblock {\em Journal of Financial Econometrics}, 7(2):174--196, 2009.

\bibitem{corsi2012discrete}
F.~Corsi and R.~Ren{\`o}.
\newblock Discrete-time volatility forecasting with persistent leverage effect
  and the link with continuous-time volatility modeling.
\newblock {\em Journal of Business \& Economic Statistics}, 30(3):368--380,
  2012.

\bibitem{dandapani2021quadratic}
A.~Dandapani, P.~Jusselin, and M.~Rosenbaum.
\newblock From quadratic hawkes processes to super-heston rough volatility
  models with zumbach effect.
\newblock {\em Quantitative Finance}, 21(8):1235--1247, 2021.

\bibitem{el2018microstructural}
O.~El~Euch, M.~Fukasawa, and M.~Rosenbaum.
\newblock The microstructural foundations of leverage effect and rough
  volatility.
\newblock {\em Finance and Stochastics}, 22(2):241--280, 2018.

\bibitem{el2019roughening}
O.~El~Euch, J.~Gatheral, and M.~Rosenbaum.
\newblock Roughening {H}eston.
\newblock {\em Risk}, pages 84--89, 2019.

\bibitem{euch2018perfect}
O.~El~Euch and M.~Rosenbaum.
\newblock Perfect hedging in rough {H}eston models.
\newblock {\em The Annals of Applied Probability}, 28(6):3813--3856, 2018.

\bibitem{el2019characteristic}
O.~El~Euch and M.~Rosenbaum.
\newblock The characteristic function of rough {H}eston models.
\newblock {\em Mathematical Finance}, 29(1):3--38, 2019.

\bibitem{filimonov2012quantifying}
V.~Filimonov and D.~Sornette.
\newblock Quantifying reflexivity in financial markets: Toward a prediction of
  flash crashes.
\newblock {\em Physical Review E}, 85(5):056108, 2012.

\bibitem{filimonov2015apparent}
V.~Filimonov and D.~Sornette.
\newblock Apparent criticality and calibration issues in the {H}awkes
  self-excited point process model: application to high-frequency financial
  data.
\newblock {\em Quantitative Finance}, 15(8):1293--1314, 2015.

\bibitem{gatheral2018volatility}
J.~Gatheral, T.~Jaisson, and M.~Rosenbaum.
\newblock Volatility is rough.
\newblock {\em Quantitative Finance}, 18(6):933--949, 2018.

\bibitem{gatheral2020quadratic}
J.~Gatheral, P.~Jusselin, and M.~Rosenbaum.
\newblock The quadratic rough {H}eston model and the joint {S\&P} 500/{VIX}
  smile calibration problem.
\newblock {\em Risk, May}, 2020.

\bibitem{goodfellow2016deep}
I.~Goodfellow, Y.~Bengio, and A.~Courville.
\newblock {\em Deep learning}.
\newblock MIT press, 2016.

\bibitem{julien2021}
J.~Guyon.
\newblock Volatility is (mostly) path-dependent.
\newblock {\em QuantMinds 2021, Barcelona}.

\bibitem{hardiman2013critical}
S.~J. Hardiman, N.~Bercot, and J.-P. Bouchaud.
\newblock Critical reflexivity in financial markets: a {H}awkes process
  analysis.
\newblock {\em The European Physical Journal B}, 86(10):1--9, 2013.

\bibitem{hochreiter1997long}
S.~Hochreiter and J.~Schmidhuber.
\newblock Long short-term memory.
\newblock {\em Neural computation}, 9(8):1735--1780, 1997.

\bibitem{jusselin2020no}
P.~Jusselin and M.~Rosenbaum.
\newblock No-arbitrage implies power-law market impact and rough volatility.
\newblock {\em Mathematical Finance}, 30(4):1309--1336, 2020.

\bibitem{patton2011volatility}
A.~J. Patton.
\newblock Volatility forecast comparison using imperfect volatility proxies.
\newblock {\em Journal of Econometrics}, 160(1):246--256, 2011.

\bibitem{patton2015good}
A.~J. Patton and K.~Sheppard.
\newblock Good volatility, bad volatility: Signed jumps and the persistence of
  volatility.
\newblock {\em Review of Economics and Statistics}, 97(3):683--697, 2015.

\bibitem{rahimikia2020machine}
E.~Rahimikia and S.-H. Poon.
\newblock Machine learning for realised volatility forecasting.
\newblock {\em Available at SSRN}, 3707796, 2020.

\bibitem{rosenbaum2021deep}
M.~Rosenbaum and J.~Zhang.
\newblock Deep calibration of the quadratic rough {H}eston model.
\newblock {\em To appear in Risk magazine}, 2021.

\bibitem{sirignano2019universal}
J.~Sirignano and R.~Cont.
\newblock Universal features of price formation in financial markets:
  perspectives from deep learning.
\newblock {\em Quantitative Finance}, 19(9):1449--1459, 2019.

\bibitem{wu2022rough}
P.~Wu, J.-F. Muzy, and E.~Bacry.
\newblock From rough to multifractal volatility: the log {S-fBM} model.
\newblock {\em arXiv preprint arXiv:2201.09516}, 2022.

\bibitem{zumbach2010volatility}
G.~Zumbach.
\newblock Volatility conditional on price trends.
\newblock {\em Quantitative Finance}, 10(4):431--442, 2010.

\end{thebibliography}
\bibliographystyle{abbrv}

\end{document}